\documentclass[epj]{svjour}
\usepackage{latexsym}
\usepackage{graphics}
\usepackage{amssymb}
\usepackage{amsmath}
\usepackage{hyperref}
\hypersetup{colorlinks=true, linkcolor=blue, urlcolor=magenta}
\begin{document}
\title{Muon Radiography to Visualise Individual Fuel Rods in Sealed Casks}
\author{Thomas Braunroth\inst{1} \and Nadine Berner\inst{2} \and Florian Rowold\inst{3} \and Marc Péridis\inst{2} \and Maik Stuke\inst{4}
}                     
%
%
\institute{thomas.braunroth@grs.de, GRS gGmbH, Schwertnergasse 1, 50667 Cologne, Germany \and nadine.berner@grs.de, marc.peridis@grs.de, GRS gGmbH, Boltzmannstr. 14, 85748 Garching, Germany \and florian.rowold@grs.de, GRS gGmbH, Kurf\"urstendamm 200, 10719 Berlin, Germany \and maik.stuke@bgz.de, BGZ mbH, Dammstra{\ss}e, 84051 Essenbach, Germany}
%
%
\abstract{
Cosmic-ray muons can be used for the non-destructive imaging of spent nuclear fuel in sealed dry storage casks.
The scattering data of the muons after traversing provides information on the thereby penetrated materials. 
Based on these properties, we investigate and discuss the theoretical feasibility of detecting single missing fuel rods in a sealed cask for the first time.
We perform simulations of a vertically standing generic cask model loaded with fuel assemblies from a pressurized water reactor and muon detectors placed above and below the cask.
By analysing the scattering angles and applying a significance ratio based on the Kolmogorov-Smirnov test statistic we conclude that missing rods can be reliably identified in a reasonable measuring time period depending on their position in the assembly and cask, and on the angular acceptance criterion of the primary, incoming muons.
%
} 
\maketitle
\label{sec:Intro}
\section{Introduction}
The operation of nuclear power plants generates high-level radioactive wastes which need to be stored and disposed of.
A well-established part of nuclear waste management is the dry storage of used fuel assemblies in designated casks. 
Depending on the availability of a final repository, the fuel assemblies might remain inside the casks placed in interim storage facilities for decades. 
The casks are designed to enclose the high-level radioactive waste and separate it from the biosphere.\newline
Inspecting the interior of a storage cask directly would require an opening of the cask, a difficult task due to the radiation.
So far, conventional radiography using neutrons or photons could not be applied successfully, partly due to the rich scattering history of traversing particles as a direct result of the dimensions of the storage cask \cite{Zio05}.
Other methods such as three-dimensional temperature field measurements or antineutrino monitoring \cite{Brdar:2017} require a detailed knowledge of the fuel history or are not suitable for the assessment of individual storage casks.  
Cosmic muons, created directly or indirectly in the atmosphere by the interaction of cosmic radiation with particles, have been also used for imaging purposes (\textit{muography}) \cite{Bon20}. 
These muons and anti-muons (muons in the following) are characterized by a broad energy spectrum spanning several orders of magnitude and have a mean momentum of approximately 4\,GeV/$c$ \cite{RPP20}.
For muons with an absolute momentum above 1\,GeV/$c$ the integral vertical intensity $I_V$ is approximately 70 m$^{-2}$s$^{-1}$sr$^{-1}$ \cite{Pas93,Gri01} and the decrease of the flux intensity scales approximately with $\cos(\theta)$, with $\theta$ being the angle with respect to the vertical \cite{RPP20}. 
In experimental physics, a value of \mbox{$I_V \approx$ 1\,cm$^{-2}$min$^{-1}$} has been established for working with horizontal detectors.\newline
Alvarez and colleagues were one of the first to use muons for non-invasive imaging and published their landmark paper on the search for hidden chambers in the pyramids of Giza in 1970 \cite{Alv70}.
These first studies were based on the measurement of the attenuation of the cosmic muon flux and provided two-dimensional projection images (\textit{muon radiography}). 
Since then, muon radiography has been used in various fields such as the study of volcanoes \cite{Tan14}, geological applications \cite{Les10}, the identification of cavities in archaeology \cite{Mor17} as well as in industrial applications \cite{Tan05}.\newline
In 2003, a Los Alamos research group proposed using the scattering angle of the outgoing muons as the basic information for imaging \cite{Bor03}.
This approach requires the measurement of the incoming as well as the outgoing trajectories of the muons and allows to obtain three-dimensional images (\textit{muon tomography}) of volumes not exceeding tens of meters.
This technique has already been used in various fields, such as nuclear control \cite{Cla15}, transport control \cite{Dec21} and the monitoring of historical buildings \cite{Zen14}.
In addition to experimental studies, Monte-Carlo simulations play an important role in muon imaging, e.g. in terms of feasibility studies or with respect to the detector design.\newline
The application of muography for the purpose of non-invasive control and monitoring of the interior of dry storage casks has gained an increasing interest and fostered experimental and simulation studies.
Besides the fundamental suitability of muography for this purpose, these efforts also addressed methodological and time requirements.
Durham \textit{et al.} \cite{Dur18} applied muon scattering radiography to a Westinghouse MC-10 cask and showed experimentally that cosmic muons can indeed be used to determine if spent fuel assemblies are missing without the need to open the cask.
A number of simulation studies were performed using two planar tracking detectors placed on opposite sides of an object, each focusing on different aspects, for example:
Jonkmans \textit{et al.} \cite{Jon13} investigated the capabilities of muons to image the contents of shielded containers to detect enclosed nuclear materials with high-$Z$.
Clarkson \textit{et al.} \cite{Cla14} performed \textsc{Geant4} simulations of a scintillating-fibre tracker for tomographic scans of legacy nuclear waste containers.
Chatzidakis \cite{Cha14} applied a Bayesian approach to monitor sealed dry casks to infer on the amount of spent nuclear fuel and to investigate the limitiations of this approach.
Using the attenuation and scattering characteristics of the muons derived from \textsc{Geant4} simulations, Ambrosino \textit{et al.} \cite{Amb15} found that a 10\,cm$^3$ Uranium block inside a concrete structure could be identified after a one-month period of measurement.
Poulsen \textit{et al.} \cite{Pou17} were the first to apply filtered back projection algorithms to muon tomography imaging of dry storage casks using simulated data and could show that this technique can be applied to the detection of missing fuel assemblies.
In a more recent work, Poulsen \textit{et al.} \cite{Pou18} used the experimental data of Ref.\,\cite{Dur18} for a numerical study using \textsc{Geant4} to distinguish different loads of a cask. 
The study indicates that a one-week muon measurement for the given experimental setup is sufficient to detect a missing fuel assembly or to identify a dummy assembly made out of iron or lead.\newline
With respect to the resolving power, both experimental and simulation studies have been focused on the level of fuel assemblies so far.
In addition, the majority of studies are based on a transversal configuration, where the detectors are placed on the sides of the cask.\newline 
In this study, we use Monte-Carlo simulations to investigate a longitudinal configuration, with the detectors placed above and below the cask.
We will investigate if muography allows detecting individual missing fuel rods. 
To unravel insights independent of reconstruction algorithms, this work will focus on radiographic images based on transmission ratios as well as scatter-angle information.\newline
The guiding questions of this work are as follows:
Is it possible to even detect individual missing fuel rods with muon radiography? 
If so, are there any constraints or requirements with respect to the experimental setup and how much time does a measurement require? 
What can be used as a significance measure to detect a missing fuel rod?
Does the significance depend on the relative position of the considered fuel rod within the fuel assembly and is the significance dependent on the number of considered events?\newline
This contribution is structured as follows:
In Sec.\,\ref{sec:Sim} we provide a detailed description of the simulation tool as well as the investigated geometry. 
Moreover, we describe the data processing and aspects related to the validation of the simulation. 
Sec.\,\ref{sec:Analysis} features the analysis and discussion. 
We address two levels of detail concentrating on the recognition of (missing) fuel assemblies and individual fuel rods.
This is followed by a summary and conclusion in Sec.\,\ref{sec:Summary}.
We end with a short outlook in Sec.\,\ref{sec:Outlook}.
\section{Simulation and Data Processing}
\label{sec:Sim}
Simulations were performed with a dedicated tool based on the Monte-Carlo toolkit \textsc{Geant4} \cite{Ago03,All06,All16}.
\textsc{Geant4} allows simulating the passage of particles through matter and has been used for numerous applications, including high energy physics, nuclear physics, accelerator physics and others.\newline
In this section we describe the key aspects of the tool, i.e. the geometry (Sec.\,\ref{ssec:Geometry}), the treatment of primary particle properties (Sec.\,\ref{ssec:PrimaryParticles}), aspects related to physics and tracking (Sec.\,\ref{ssec:PhysicsTracking}) as well as optimization strategies (Sec.\,\ref{ssec:Optimizations}) to reduce the computational time.\newline
The tool was compiled against v10.06.p2 of \textsc{Geant4} and allows using multithreading.
The results are written event-by-event into \textsc{ROOT} \cite{Bru97,ROOT20} container files, which allows performing post-processing in a flexible manner.
Finally, Sec.\,\ref{ssec:Validation} discusses the validation of the code by comparing simulated and tabulated (or empirically established) energy losses and angular straggling for different target materials and projectile energies.
\subsection{Geometry}
\label{ssec:Geometry}
\subsubsection{Generic Cask}
\label{sssec:GenericCask}
\begin{figure}[t]
\resizebox{1.0\columnwidth}{!}{\includegraphics{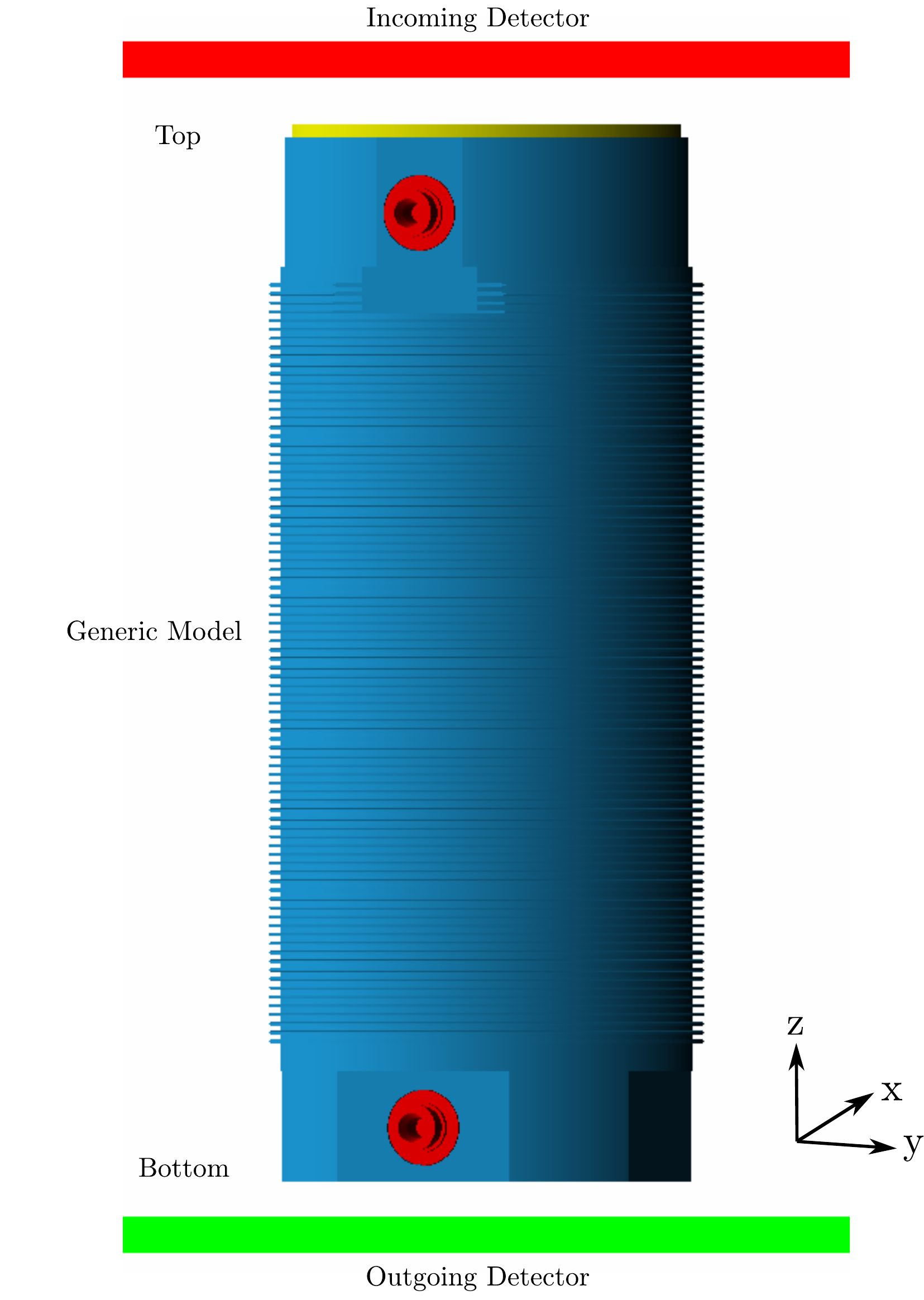}}
\caption{Visualisation of the generic model and its upward orientation within the present work. The coordinate system as it is used in this study is provided in the lower right corner. The origin of the coordinate system coincides with the center position of the bottom part of the model. The red and green areas above and below the model indicate the incoming and outgoing detectors.}
\label{Fig:FullGenModel}
\end{figure}

\begin{figure*}[t]
\resizebox{1.0\textwidth}{!}{\includegraphics{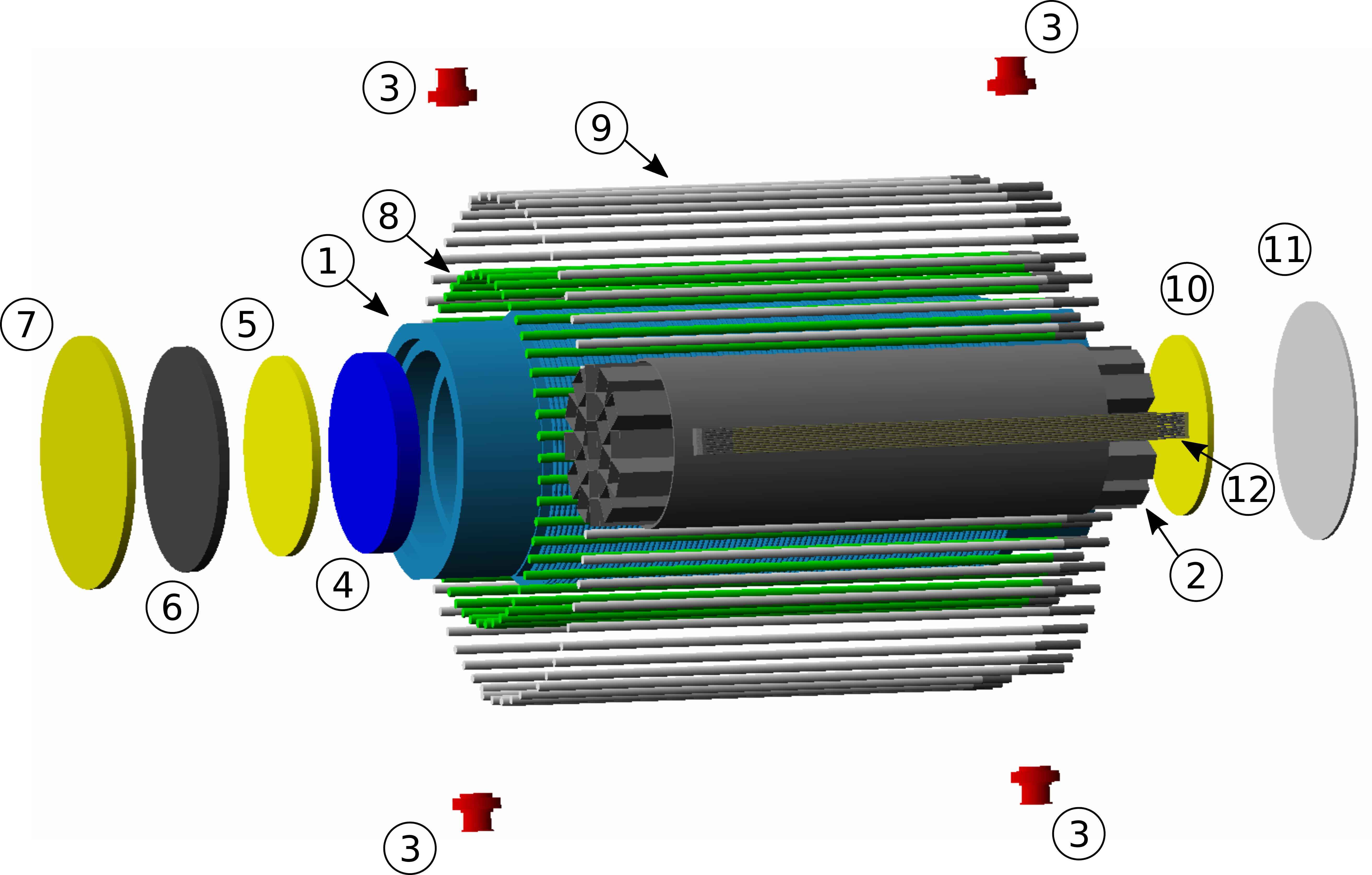}}
\caption{Exploded view of the generic model showing its individual components: Monolithic body (1), basket with 19 fuel compartments (2), trunnions (3), primary lid (4), polyethylene plate (5), secondary lid (6), protection plate (7), inner moderator rods and plugs (8), outer moderator rods and plugs (9), polyethylene plate (10), base plate (11) and a representative fuel assembly (12). See text for details.}
\label{Fig:GenModelExplodedView}
\end{figure*}

The key component of the geometry is a generic cask model (referred to as generic model in the following) which mimics the features of the CASTOR\textsuperscript{\textregistered} V/19 cask \cite{GNS20}, e.g. in terms of major components, dimensions, materials and masses.
The CASTOR\textsuperscript{\textregistered} V/19 cask is used for transport as well as storage purposes and is designed to carry up to 19 fuel assemblies from pressurized water reactors (PWR).
All information on geometries and materials specifying the generic model was derived from public available sources such as Refs.\,\cite{GNS20,BFS00}.\newline
A visualisation of the generic model was generated with the \textsc{Geant4} OpenGL interface and is depicted in Fig.\,\ref{Fig:FullGenModel}.
The individual components of the generic model can be identified in the exploded view shown in Fig.\,\ref{Fig:GenModelExplodedView}.\newline
A comparison of some key properties of the generic model on the one hand and the CASTOR\textsuperscript{\textregistered} V/19 cask on the other hand can be found in Table\,\ref{Tab:CompGenModelVsCV19} and highlights the mutual similarities.\newline
Each of the 19 fuel compartments can be occupied with one (modelled) fuel assembly.
Each modelled fuel assembly comprises top- and bottom-nozzle, fuel rods as well as control rods. 
%
The fuel rods consist of nuclear fuel (UO$_2$) and cladding tubes (Zirconium alloy). 
A complete 18x18-24 fuel assembly consists of 300 fuel rods and 24 control rods.
The arrangement of fuel and control rods within a complete fuel assembly is shown in Fig.\,\ref{Fig:RodArrangement}.
The modelled top and base components of the assembly are simplified.
They are assumed to be box-like, with heights chosen to comply with real masses.
Basic properties related to the modelled fuel assemblies are summarized in Table\,\ref{Tab:PropertiesFuelElement}.\newline
\begin{table}[t]
\caption{Comparison of key dimensions between the generic model and the CASTOR\textsuperscript{\textregistered} V/19 cask (in storage configuration).}
\label{Tab:CompGenModelVsCV19}       
\begin{tabular}{lcc}
\hline\noalign{\smallskip}
Property & Generic Model & CASTOR\textsuperscript{\textregistered}\\
 & Model & V/19 \cite{GNS20}\\
\noalign{\smallskip}\hline\noalign{\smallskip}
Overall Height & 594\,cm & 594\,cm \\
Outer Diameter & 244\,cm & 244\,cm \\
Cavity Height & 502.5\,cm & 503\,cm \\
Cavity Diameter & 148\,cm & 148\,cm \\
\noalign{\smallskip}\hline
\end{tabular}
\end{table}
All parts of the \textsc{Geant4} model were derived from basic \textsc{Geant4} solid objects and refined with Boolean operations. 
Each component can be switched on and off by a command-based user interface, which easily allows performing simulations for different target geometries.
In addition, it is possible to remove arbitrary components from a fuel assembly, e.g. specific fuel rods at specific slot positions.
This allows, among others, for investigating the contribution of specific fuel rods to radiographic (or tomographic) images in more detail. 
\subsubsection{Box-Like Object}
\label{sssec:BoxLikeObject}
Instead of the generic model, it is possible to generate a box-like object, whose basic properties - i.e. dimensions, placement and material - can be specified individually for each simulation run.
All interactions within this box are recorded within \textsc{ROOT} container files, particularly useful for validation purposes.
\begin{table}[b]
\begin{center}
\caption{Properties of the modelled fuel assemblies.}
\label{Tab:PropertiesFuelElement}       
\begin{tabular}{lcc}
\hline\noalign{\smallskip}
Value/Parameter & Property\\
\noalign{\smallskip}\hline\noalign{\smallskip}
Material - Head and Base  & Stainless Steel \\
Material - Cladding &  Zirconium Alloy\\
Material - Fuel & Uranium Dioxide\\
Material - Control Rods & Zirconium Alloy\\
Length - Fuel Rod & 4407\,mm\\
Length - Active Length & 3900\,mm\\
Number of Control Rods &  24\\
Number of Fuel Rods & 300 \\
Total Weight & 845\,kg\\
\noalign{\smallskip}\hline
\end{tabular}
\end{center}
\end{table}
\subsubsection{Coordinate System}
\label{sssec:CoordinateSystem}
The $z$-axis coincides with the symmetry axis of the generic model and is oriented upwards, i.e. its orientation is selected so that the $z$-coordinate of the model's top is larger than the $z$-coordinate of its bottom, $z_{\text{top}} > z_{\text{bottom}}$.
The $x$ and $y$ axis are orientated in such a way that the $(x,y,z)$-coordinate system generates a right-handed euclidean space.
The orientation of the individual axes is indicated in Fig.\,\ref{Fig:FullGenModel}.
The angle $\theta$ reflects the angle of a given vector $\vec{r}$ and the inverted $z$-axis. 
Using this convention, a muon from the zenith is characterized by $\theta = 0^{\circ}$.
The angle $\varphi$ is the angle between the projection of a vector $\vec{r}$ onto the $(x,y)$-plane and $\hat{e}_x$.
\subsubsection{Detectors}
\label{sssec:Detectors}
Detector systems are mimicked by two rectangular detector planes (vanishing thickness, area of $(3\times 3)\,\text{m}^2$), whose normal vectors are parallel to $\hat{e}_z$. 
The detector plane placed above the generic model is called incoming detector whereas the detector plane placed below the generic model is called outgoing detector.
The gap between the detector surface and the top (or the bottom) of the generic cask is $\approx 10$\,cm.
Muons traversing these planes are tracked and key information is determined, see Sec.\,\ref{ssec:PhysicsTracking}.\newline
This approach gives access to at least as many properties as a real detection system for muon tomography may provide.
\begin{figure}[b]
\resizebox{1.00\columnwidth}{!}{\includegraphics{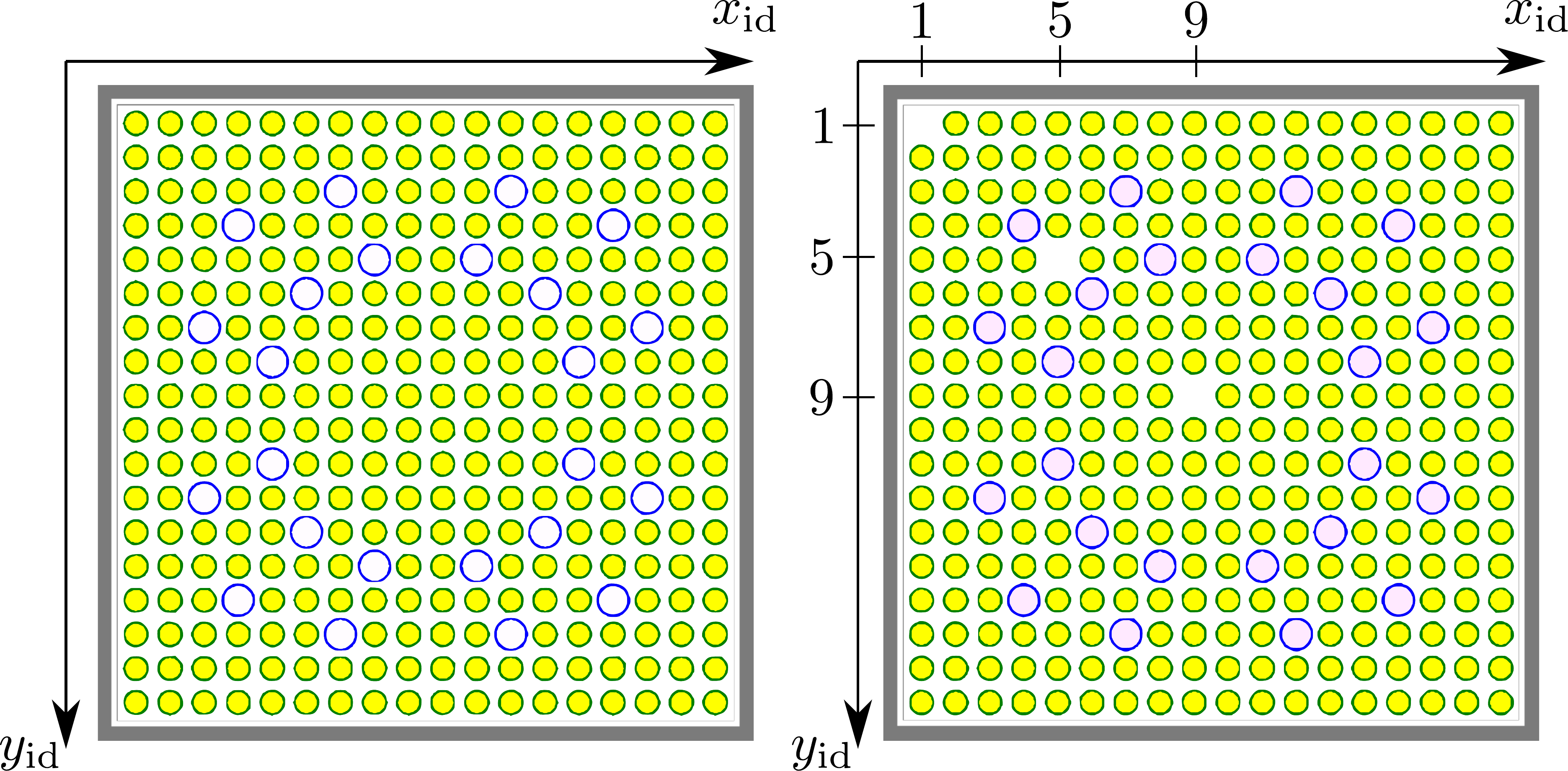}}
\caption{\textit{Left}: Top-view on the arrangement of fuel rods (yellow), cladding tubes (green) as well as control rods (blue) in case of a complete fuel assembly. Each element can be specified unambiguously based on its slot position given by $x_{\text{id}}$, $y_{\text{id}}$. \textit{Right}: Same as left figure, but several elements along the diagonal are missing ($x_{\text{id}}/y_{\text{id}} = 1, 5, 9$). This configuration is investigated in more detail in Sec.\,\ref{SSec:AnaSingleFuelElement}.}
\label{Fig:RodArrangement}
\end{figure}
Since the simulation allows determining all properties precisely on an event-by-event base, it benefits from an infinite resolving power\footnote{In principle, the present approach allows to include resolution effects in the post-processing without the need to repeat time-consuming simulations.}.
In this regard, it provides a best-case scenario and can be considered as a first important theoretical step towards substantiated feasibility studies.
\subsection{Primary Particles}
\label{ssec:PrimaryParticles}
Primaries are generated by a user-defined class based on the G4VUserPrimaryGeneratorAction class provided by \textsc{Geant4}.\newline
The Monte-Carlo approach is realized by using uniform random number generators with limits $a$, $b$ ($\mathcal{U}[a,b]$) to determine for each event $k$ the particle properties, i.e. particle type (muon ($\mu^-$) or anti-muon ($\mu^+$)), initial momentum information ($p_{\mu_k}$, $\theta_k$, $\varphi_k$) as well as initial position information ($x_k$, $y_k$).
Details are described in the following.
\subsubsection*{Particle Type}
All primaries are muons or anti-muons, assuming a charge ratio $\mu^+/\mu^-$ equal to 1.28 \cite{CMS10}.
The particle type for the event $k$ is determined assuming $\mathcal{U}[0, 2.28]$ - if the generated random number is smaller than (or equal to) 1.28, the generated primary for this event will be an anti-muon ($\mu^+$), otherwise it will be a muon ($\mu^-$).
\subsubsection*{Initial Particle Momentum}
\begin{figure}[t]
\resizebox{1.0\columnwidth}{!}{\includegraphics{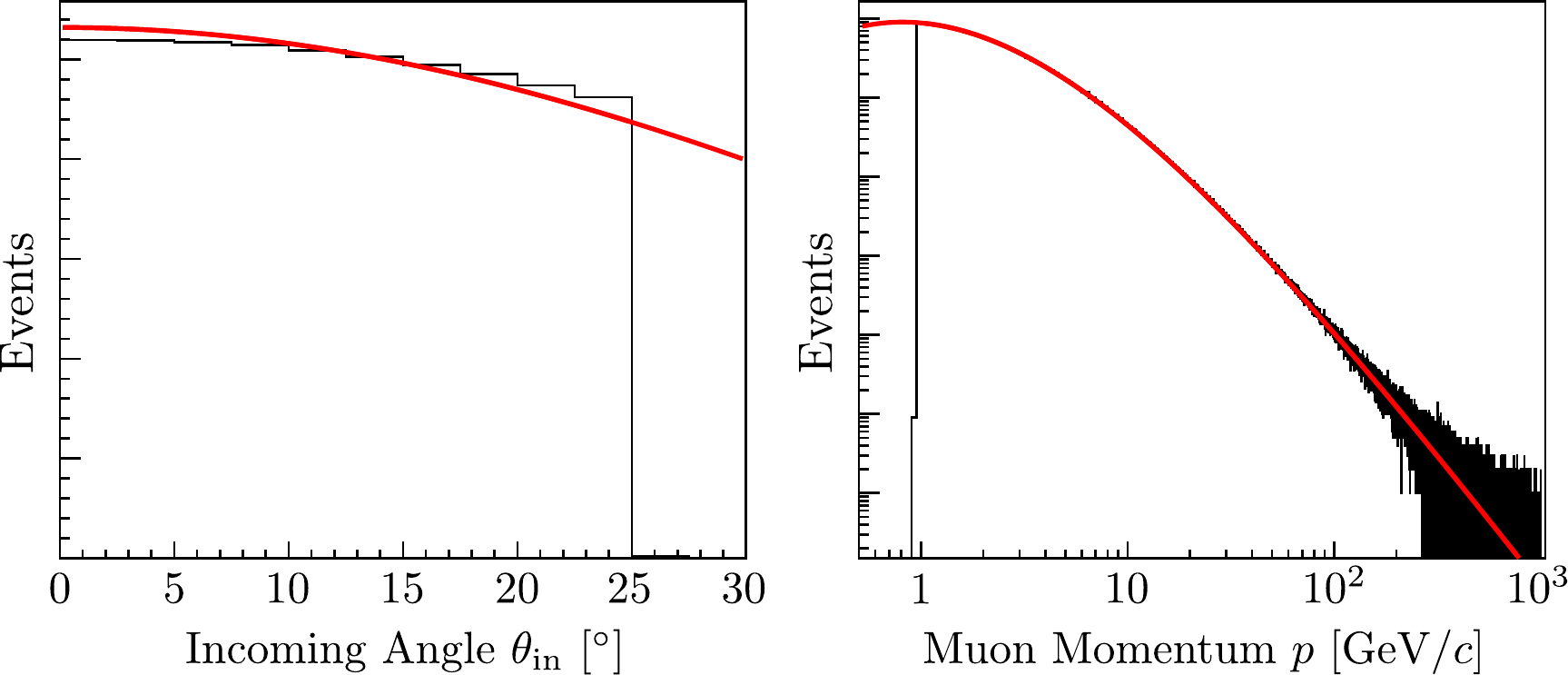}}
\caption{Sample distributions of the initial angle $\theta_{\text{in}}$ (\textit{left}) and of the absolute muon momentum $p$ (\textit{right}) as determined with the incoming detector. The red curve shown in the left spectrum is proportional to $\cos^2\left(\theta\right)$, while the red curve shown in the right spectrum is proportional to Eq.\,\ref{Eq:Reyna} with $\theta = 12.5^{\circ}$.}
\label{Fig:AngleAndMomentum}
\end{figure}

The description of the initial muon momentum is based on an empirical and parametric approach \cite{Rey06}, according to which the muon intensity at any given momentum and angle to the vertical intensity $I_V$ is given by
\begin{equation}
I(p_{\mu}, \theta) = \cos^3(\theta)\cdot I_V(\zeta),\quad \zeta = p_{\mu}\cdot\cos\left(\theta\right)
\label{Eq:Reyna}
\end{equation}
Here, $I_V$ is given by the Bugaev parametrisation \cite{Bug98}:
\begin{equation}
I_V(p_{\mu}) = c_1\cdot p_{\mu}^{-(c_2 + c_3\cdot\log_{10}\left( p_{\mu}\right) + c_4\cdot\log^2_{10}\left( p_{\mu}\right) + c_5\cdot\log^3_{10}\left( p_{\mu}\right))}
\label{Eq:Bug}
\end{equation} 
The parameters $c_1$ to $c_5$ were determined in Ref.\,\cite{Rey06} by a fitting approach and are quantified as:
\begin{eqnarray*}
c_1 &=& 0.00253\\
c_2 &=& 0.2455\\
c_3 &=& 1.288\\
c_4 &=& -0.2555\\
c_5 &=& 0.0209
\end{eqnarray*}
Due to the lack of a suitable random number generator to directly address the associated probability distribution in \textsc{Geant4}, the analytical description is discretized in a two-dimensional pattern as follows.\newline
Firstly, the polar component of the angular spectrum represented by $\theta_{\text{in}}$, ranging from $\theta_{\text{min}} = 0^{\circ}$ to an upper limit of $\theta_{\text{max}} = 25^{\circ}$, is split into $n$ bins $b_{\theta,i}$ ($ i=1,...,n$).
The quoted upper limit of $25^{\circ}$ limits the contributions of trajectories that geometrically can only be detected by one of the detectors. 
All $b_{\theta,i}$ span over identical angular ranges.
For each angular bin $b_{\theta,i}$, the lower and upper limits are then given by $\theta^{\text{ll}}_{i}[^{\circ}] = 25\,\frac{(i-1)}{n}$ and $\theta^{\text{ul}}_{i}[^{\circ}] = 25\,\frac{i}{n}$.\newline
The function $I_i(p_{\mu},\overline{\theta}_i)$, for $i = 1,..,n$, can then be associated to each of these bins of the angular spectrum, where $\overline{\theta}_i$ is the arithmetic mean angle of the specific bin.
These assigned functions $I_i$ are then integrated numerically within the kinetic energy range of $T \in [1\,\text{GeV}, 1\,\text{TeV}]$.
The calculated integrals quantify the relative weights $w_i$ for each of the angular bins $b_{\theta,i}$:
\begin{equation}
w_i = \int_{p_{\mu}(T=1\,\text{GeV})}^{p_{\mu}(T=1\,\text{TeV})}{dp'_{\mu}\,I_i(p'_{\mu},\overline{\theta}_i)}
\end{equation}
Information on $\theta^{\text{ll}}_{i}$, $\theta^{\text{ul}}_{i}$ and $w_i$ are stored in a dedicated text file that is used as an input parameter to the simulation.
$\mathcal{U}[0,\,\sum_i{w_i}]$ uses these weights $w_i$ to determine for each event $k$ the proper angular bin $b_{\theta,k}$.
In a next step, the proper polar angle $\theta_k$ for the present event $k$ is determined using $\mathcal{U}[\theta^{\text{ll}}_{k},\theta^{\text{ul}}_{k}]$.
The azimuthal angle $\varphi_k$ is determined for each event $k$ using $\mathcal{U}[-\pi, \pi]$.\newline
Secondly, the absolute momentum spectra are treated comparably.
The momentum axis from $p_{\mu}(T=1\,$GeV) to $p_{\mu}(T=1\,$TeV) is discretized into $m$ bins $b_{p_{\mu},j}$ ($j = 1,...,m$) with increasing bin sizes.
Each bin $b_{p_{\mu},j}$ is characterized by the following integral:
\begin{equation}
v_{i,j} = \int_{p^{\text{ll}}_{\mu,j}}^{p^{\text{ul}}_{\mu,j}}{dp'_{\mu}\,I_i(p'_{\mu},\overline{\theta}_i)}
\end{equation}
Here, $p^{\text{ll}}_{\mu,j}$ and $p^{\text{ul}}_{\mu,j}$ denote the lower and upper limits of the bin $b_{p_{\mu},j}$.
\newline
For each angular bin $b_{\theta, i}$, values for $p^{\text{ll}}_{\mu,j}$, $p^{\text{ul}}_{\mu,j}$ and $v_{i,j}$ are stored in dedicated text-files which are used as mandatory input information for the simulation code.\newline
$\mathcal{U}[0,\,\sum_j{v_{i,j}}]$ uses these weights $v_{i,j}$ to determine for each event $k$ the proper momentum bin $b_{p_{\mu},k}$.
Finally, the proper $p_{\mu,k}$ is determined using $\mathcal{U}[p^{\text{ll}}_{\mu,k},\, p^{\text{ul}}_{\mu,k}]$.
Within the present work, all numerical integrations were performed with \textsc{GNU Octave} \cite{Eat19}.\newline
Fig.\,\ref{Fig:AngleAndMomentum} shows histograms of the simulated angular and  momentum distributions for $\theta_{\text{min}}=0^{\circ}$, $\theta_{\text{max}}=25^{\circ}$, $n=10$ and $m=172$.\newline
In addition to this distribution-based approach described above, the tool also allows performing simulations with mono-energetic and mono-directional muons.
\subsubsection*{Initial Particle Position}
The initial positions of the muons are selected as follows.
The $z$-coordinate is a fixed value and selected so that it is ensured that the muon is created just above the incoming detector.
The $x$- and $y$-coordinates are selected from uniform distributions covering ranges from $x_{\text{min}}$ to $x_{\text{max}}$ ($\mathcal{U}[x_{\text{min}},x_{\text{max}}]$) and $y_{\text{min}}$ to $y_{\text{max}}$ ($\mathcal{U}[y_{\text{min}},y_{\text{max}}]$), respectively.
The associated limits can be specified in the input file of the simulation.
\subsection{Physics and Tracking}
\label{ssec:PhysicsTracking}
This code uses the modular physics list FTFP{\_}BERT \linebreak implemented in \textsc{Geant4}, which is recommended by the \textsc{Geant4} developers for high-energy physics.\\
The incoming and outgoing detector volumes are linked to a dedicated tracking class.
This allows tracking and storing information on the particle properties needed for the post processing to generate radiographic (or tomographic) images.
\subsection{Cutting Conditions to Reduce Computational Time}
\label{ssec:Optimizations}
Muons may interact with the material of the generic model according to the chosen physics list in various ways.
Some of these interactions are irrelevant to the present work.
For example, some interactions may generate secondary particles - e.g. photons or neutrons - whose tracking consumes computational time without any benefits or consequences to the generated images.
To avoid an unnecessary increase of the computational time, the trajectories of such secondary particles are annihilated at the end of their first steps. 
It must be stressed that this does not concern the possible interactions of muons with matter - all interactions which may lead to such secondary particles still take place.\newline
In addition and as already described in Sec.\,\ref{ssec:PrimaryParticles}, the muon properties - with respect to the angle $\theta$ and kinetic energy $T$ - were cut to avoid the simulation of muon trajectories that are not useful for an analysis based on a two-detector setup. 
Finally, all muon trajectories are annihilated at the exit of the outgoing detector.
\subsection{Validation}
\label{ssec:Validation}
This section discusses the validation of two key aspects with respect to muon imaging. 
The first aspect deals with the energy loss of muons in matter and is discussed in Sec.\,\ref{SSSec:ELoss}.
The second aspect deals with the angular scattering of muons after traversing matter with a known thickness Sec.\,\ref{SSSec:AngScatt}.
In both cases, the referenced thickness was specified to 1\,mm.
\subsubsection{Energy Loss of Muons In Various Target Materials}
\label{SSSec:ELoss}
\begin{figure}[htb]
\resizebox{1.0\columnwidth}{!}{\includegraphics{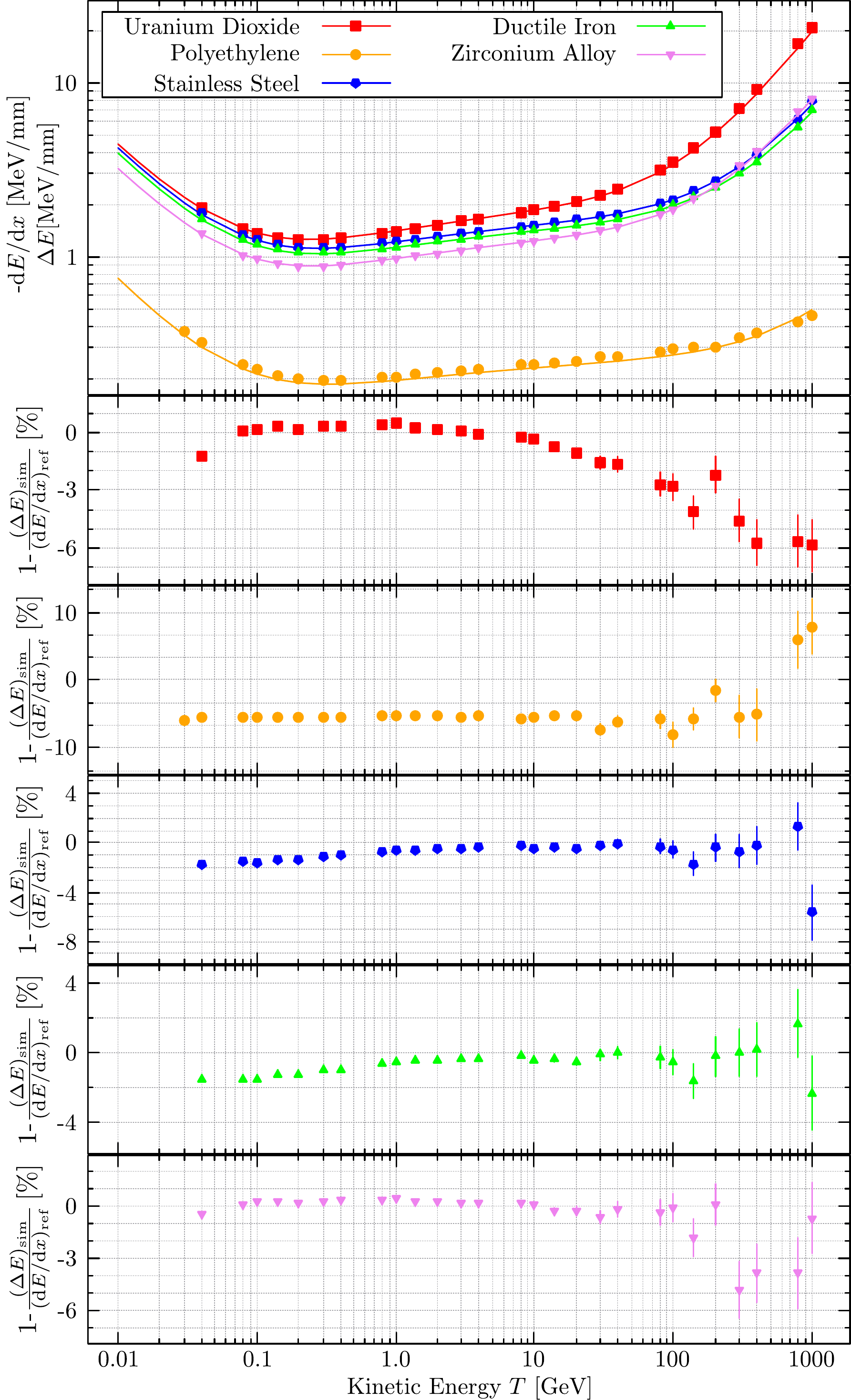}}
\caption{Comparison of simulated mean energy losses $\Delta E$ per distance and mean differential energy losses d$E$/d$x$ (both in [MeV/mm]) of muons in various target materials. The top figure shows simulated mean energy losses (symbols) and compares them to tabulated reference values (solid lines). The red curve (boxes) shows the results for uranium dioxide, the orange curve (circles) corresponds to polyethylene, the blue curve (pentagons) corresponds to stainless steel, the green curve (up-pointing triangles) corresponds to ductile iron and the violet curve (down-pointing triangles) corresponds to zirconium alloy. The lower five figures show for each material the relative deviations between simulated energy loss per mm and tabulated mean stopping power values.}
\label{Fig:ValdEdx}
\end{figure}

In this section, we compare simulated energy losses of muons in relevant target materials - uranium dioxide, poly{\-}ethylene, stainless steel, ductile iron and zirconium alloy - to tabulated (or calculated) values of the mean differential energy loss \cite{Gro01,ANP20}.\newline
The quoted references provide only for two of the listed compound materials - uranium dioxide and polyethylene - tabulated values. 
Hence, reference values were calculated for the other compounds - stainless steel, ductile iron and zirconium alloy - according to Bragg's Rule of Stopping Power Additivity \cite{Bra05}:
\begin{eqnarray}
w_j &=& \frac{n_j\cdot A_j}{\sum_k{n_k\cdot A_k}} \\
\frac{\text{d}E}{\text{d}x} &=& \sum_j{w_j\cdot\left.{\frac{\text{d}E}{\text{d}x}}\right\vert_j}
\end{eqnarray}
Here, $w_j$ denotes the mass fraction of the material $j$, d$E$/d$x\vert_j$ is the differential energy loss in the material $j$ and d$E$/d$x$ describes the mean differential energy loss by the muons in the compound material.\newline
Strictly speaking, the simulated energy losses per distance $\Delta E$ are not identical to differential energy losses.
To increase the comparability, we consider for both numbers a distance of 1\,mm and cut the lower limit of the considered kinetic energy range to ensure that the (simulated) mean energy losses are smaller (or comparable) to five percent of the kinetic energy $T$, i.e. $\Delta E \lesssim 0.05\,T$.
The upper limit of the kinetic energy range is set to $T=1$\,TeV which equals the maximal kinetic energy of the considered muon primaries in the present study.\newline
The number of simulated events was chosen in a way that the relative uncertainties of the mean values $\Delta E$ extracted from the simulated energy-spectra are below 2\,\%.
The only exception is given for polyethylene, for which the limit is increased to 5\,\%.
In total, between $10^5$ and $5\cdot 10^6$ events were simulated for each kinetic energy and each material.
The results are shown in Fig.\,\ref{Fig:ValdEdx}.\newline
For the materials stainless steel, ductile iron and zirconium alloy, the agreement between simulated and tabulated values is very good and the majority of values deviate by less than 2\,\%.
Few exceptions were found for kinetic energies close to 1\,GeV.
It shall be mentioned that due to the high $\Delta E$ dispersion at high projectile energies, the mean values are quite sensitive to statistical outliers.\newline
The largest deviations were found for polyethylene, for which relative deviations close to 7\,\% were determined over a broad kinetic energy range.\newline
For uranium dioxide, the agreement between $T=80$\,MeV and $T=10$\,GeV is very good.
However, with increasing kinetic energies a clear trend towards larger deviations of up to $\sim 6$\,\% was observed.\newline
In summary, the overall agreement between simulated and tabulated (or calculated) stopping power values is convincing. 
In general and with respect to the given kinetic energy limits, the simulation appears to slightly overestimate the stopping power of muons in the investigated materials.
\subsubsection{Angular Scattering of Muons in Various Target Materials}
\label{SSSec:AngScatt}
In this section, we investigate simulated scattering angles $\theta^S$ of muons after traversing different target materials and compare the width of the associated distributions to an established semi-empirical description.
Each material had a thickness of 1\,mm.\newline
According to Ref.\,\cite{Lyn91}, the root-mean square of the scattering angle $\theta^S_{p_{x,y}}$ in the $x$- and $y$-planes can be described using the following formula:
\begin{equation}
\mbox{
\begin{small}
$\sigma_{\theta^S_{p_{x,y}}} = \frac{13.6\,\text{MeV}}{\beta\cdot p\cdot c} \sqrt{\frac{\Delta z}{X_0}}\cdot\left.\left[{1+ 0.038\cdot\ln{\left.\left({\frac{\Delta z}{X_0\cdot \beta^2}}\right.\right)}}\right.\right]
$
\end{small}
}
\end{equation}
Here, $\beta$ is given by $v/c$, $p$ is the absolute muon momentum, $\Delta z$ is the thickness of the irradiated material and $X_0$ is the radiation length of the material.
\begin{figure}[hbt]
\resizebox{1.0\columnwidth}{!}{\includegraphics{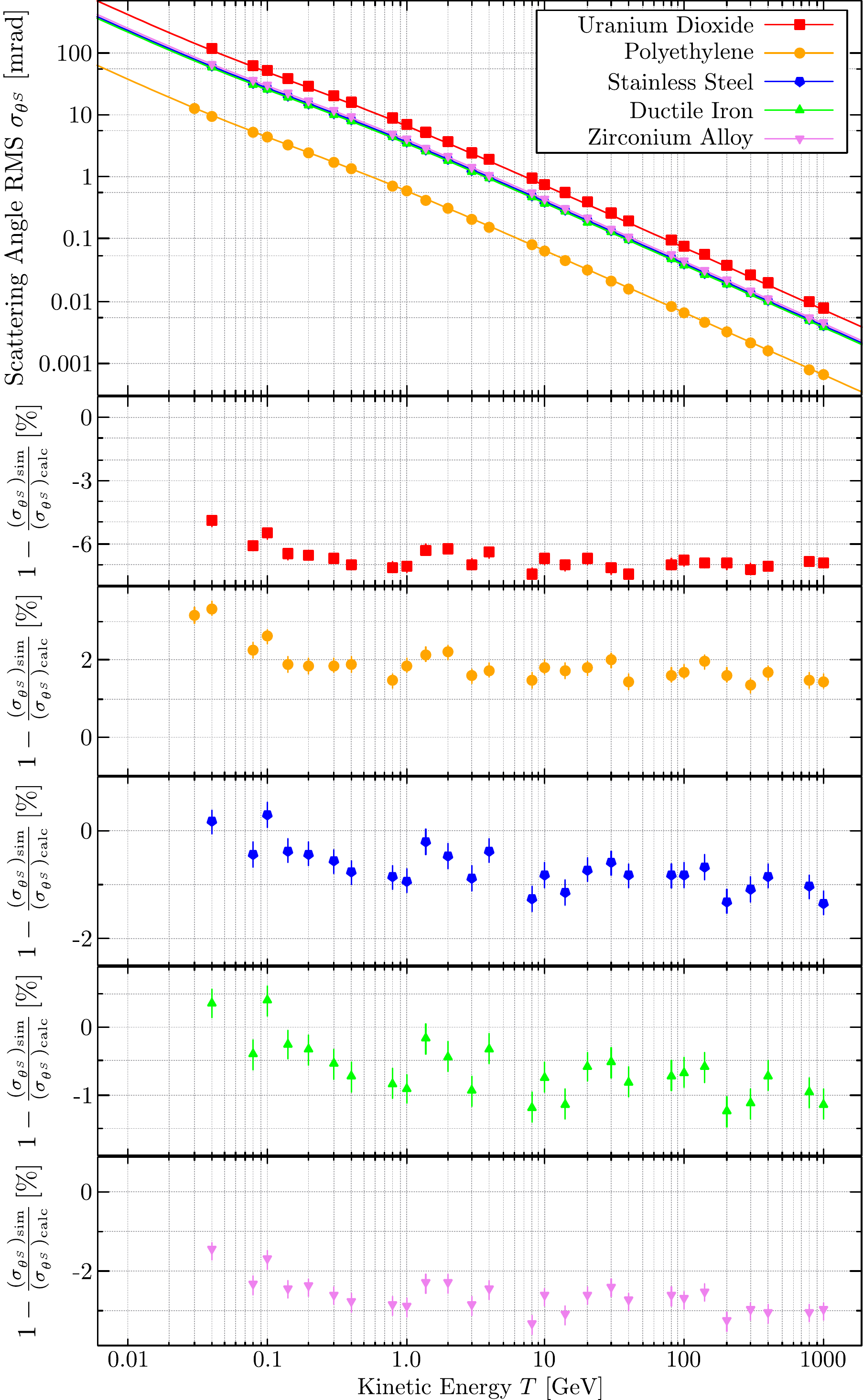}}
\caption{Comparison of root-mean square (RMS) values of scattering angles (in [mrad]) between semi-empirical calculations and simulated values for muons in various target materials with a thickness of 1\,mm. The top figure shows values derived from the simulation (symbols) and compares them to calculated values based on a semi-empirical approach (solid lines). The red curve (boxes) shows the results for uranium dioxide, the orange curve (circles) corresponds to polyethylene, the blue curve (pentagons) corresponds to stainless steel, the green curve (up-pointing triangles) corresponds to ductile iron and the violet curve (down-pointing triangles) corresponds to zirconium alloy. The lower five figures show for each material the relative deviations between simulated and calculated values.}
\label{Fig:ValScattering}
\end{figure}
\begin{figure*}[t]
\resizebox{1.0\textwidth}{!}{\includegraphics{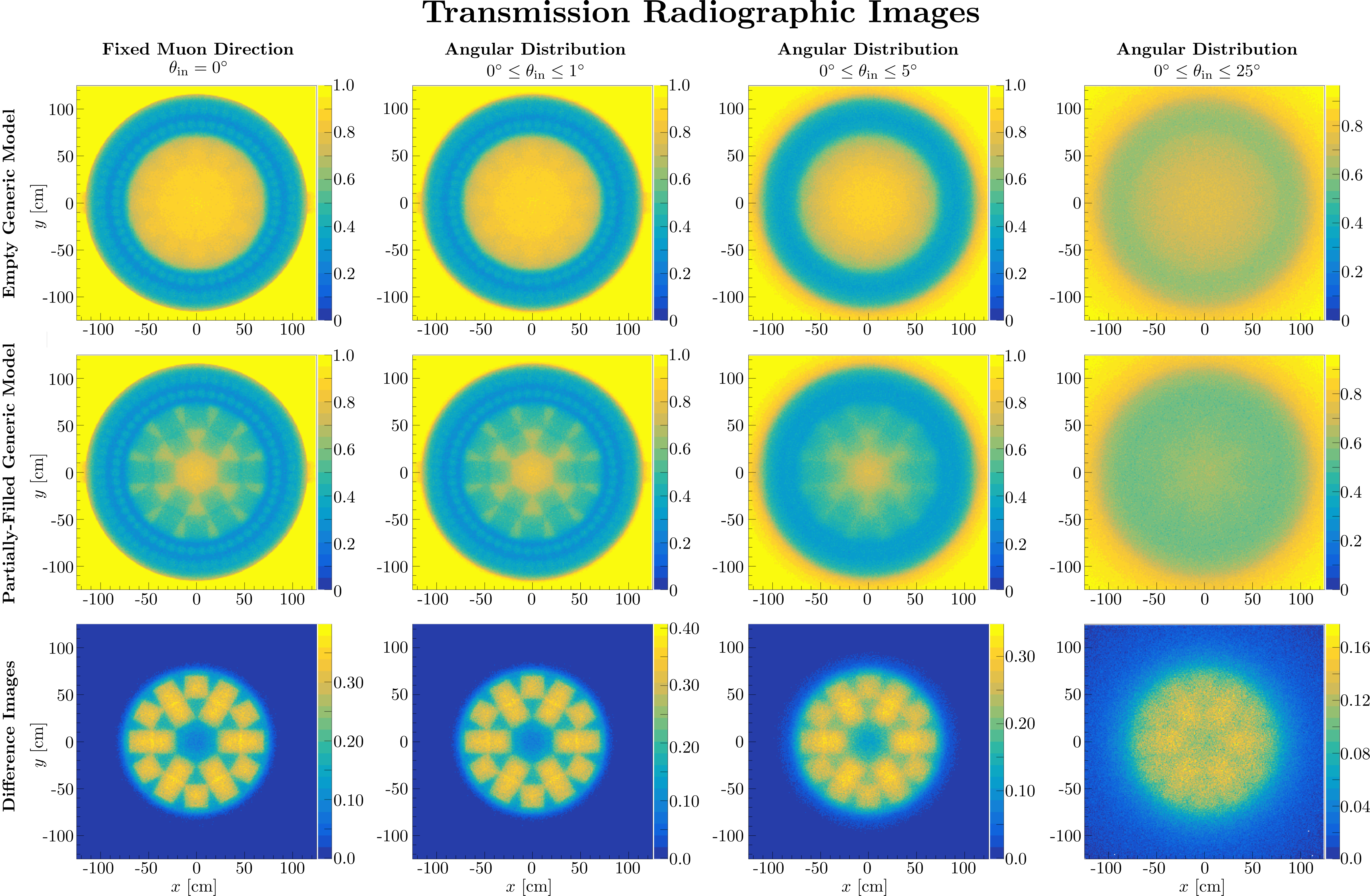}}
\caption{The figures in the two upper rows show transmission radiographic images of longitudinal scans of $a$.) an empty generic model (\textit{top row}) as well as $b$.) a generic model filled with 18 out of 19 possible fuel assemblies (\textit{middle row}) where the central fuel compartment remained empty. The \textit{bottom row} shows corresponding difference images, $a-b$. The color code represents the ratio of muons that were detected in both detectors over the total number of muons. Different assumptions on the initial muon directions are considered from left to right and range from a fixed initial muon direction with $\theta_{\text{in}} = 0^{\circ}$ (\textit{left}) to an angular distribution with respect to $\theta_{\text{in}}$ of $0^{\circ}$ to $25^{\circ}$ (\textit{right}). The $x$- and $y$-coordinates refer to the muon positions at the exit of the incoming detector.}
\label{Fig:IntView_RadPlots}
\end{figure*}
The root-mean square of the scattering angle $\theta^S$ in space is given by:
\begin{equation}
\sigma_{\theta^S} = \sqrt{2}\,\sigma_{\theta^S_{p_{x,y}}}
\end{equation}
In case of a compound material, the radiation length $X_0$ can be calculated using the formula
\begin{equation}
\frac{1}{X_0} = \sum_j{\frac{w_j}{X_{0,j}}}
\label{Eq:X0}
\end{equation}
where $w_j$ is the mass fraction of the component $j$ and $X_{0,j}$ is the radiation length of the component $j$.
\begin{table}[b]
\begin{center}
\caption{Radiation lengths $X_0$ for materials relevant to the present work.}
\label{Tab:RadLength}       
\begin{tabular}{lcc}
\hline\noalign{\smallskip}
Material & $X_0$\,$[\text{g/cm}^2]$ & $X_0$\,[cm]\\
\noalign{\smallskip}\hline\noalign{\smallskip}
Uranium Oxide & 6.65 & 0.6068\\
Polyethylene & 44.77 & 50.31\\
Stainless Steel & 13.921 & 1.808\\
Ductile Iron & 14.297 & 2.014\\
Zirconium Alloy & 10.223 & 1.558\\
\noalign{\smallskip}\hline
\end{tabular}
\end{center}
\end{table}
The radiation lengths were either taken directly from Refs.\,\cite{Gro01,ANP20} or calculated according to Eq.\,\ref{Eq:X0}.
An overview of the various radiation lengths $X_0$ for the relevant materials is given in Table\,\ref{Tab:RadLength}.\newline
Simulations covering a broad range of kinetic energies were performed for all relevant materials with $10^5$ events.
With respect to the energy range, the same restrictions as for the discussion of the energy loss were applied.
The simulated results are shown and compared to calculated values in Fig.\,\ref{Fig:ValScattering}.\newline
The deviations between simulated and semi-empirical $\sigma_{\theta^S}$ are usually less than 3\,\%.
Only for UO$_2$ larger deviations of about 6\,\% occur.
In all cases, the deviations as a function of the kinetic energy remain rather constant.
Overall, the agreement between simulated and calculated values is very good.
\section{Analysis and Discussion}
\label{sec:Analysis}
\begin{figure*}[!t]
\resizebox{1.0\textwidth}{!}{\includegraphics{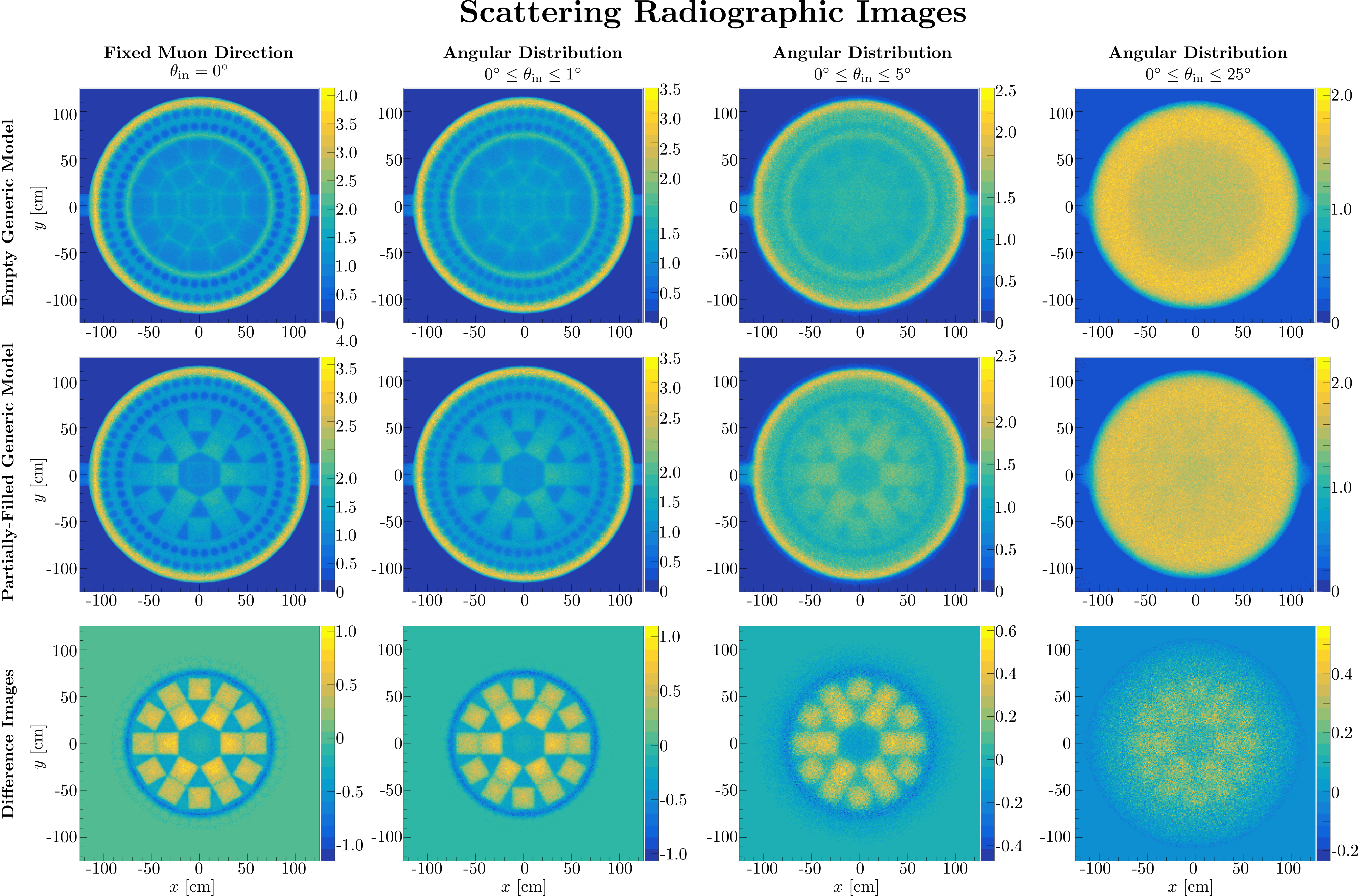}}
\caption{The figures in the two upper rows show scattering transmission images of longitudinal scans of $a$.) an empty generic model (\textit{top row}) as well as $b$.) a generic model filled with 18 out of 19 possible fuel assemblies (\textit{middle row}) where the central fuel compartment remained empty. The color code represents for each pixel the median of the associated scattering angle distribution. The \textit{bottom row} shows corresponding difference images, $b-a$. Different assumptions on the initial muon directions are considered from left to right and range from a fixed initial muon direction with $\theta_{\text{in}} = 0^{\circ}$ (\textit{left}) to an angular distribution with respect to $\theta_{\text{in}}$ of $0^{\circ}$ to $25^{\circ}$ (\textit{right}). The $x$- and $y$-coordinates refer to the muon positions at the exit of the incoming detector.}
\label{Fig:IntView_TomPlots}
\end{figure*}
The analysis is split into two main sections.
The first section investigates longitudinal scans of the ge{\-}neric model and analyses simulated radiographic images.
Here, the focus lies on the effects of various angular acceptances on the image quality.
In addition, we compare transmission radiographic images with scattering radiographic images, e.g. in terms of image contrast and resolving power.
The presentation focuses on qualitative aspects and is described in more detail in Sec.\,\ref{Ssec:OverallScan}.\newline
The second section provides a detailed study that focuses on the central fuel assembly.
In particular, it investigates the suitability of muon scattering radiography to make statements about the occupancy of individual fuel-rod slots within a certain fuel assembly.
This section focuses on quantitative aspects and is described in Sec.\,\ref{SSec:AnaSingleFuelElement} in more detail. 
\subsection{Longitudinal Scans Covering the Full Cross Section of the Generic Model}
\label{Ssec:OverallScan}
All simulations discussed in this section were performed with $5\cdot 10^7$ events.
All primary muons were generated at a fixed height ($z\approx 6.1$\,m) just above the generic model, while the initial $x$- and $y$-coordinates followed uniform distributions within the limits of $-1.25 \,{\text{m}} \leq x,y \leq 1.25 \,{\text{m}}$.
This area of the initial flux is sufficient to cover the full cross-section surface of the generic model.\newline
The absolute muon momenta $p$ were treated as described in Sec\,\ref{ssec:PrimaryParticles}, while different scenarios were simulated with respect to the initial muon direction.
As a first scenario (reference case), we assumed mono-directional initial muons, i.e. $\vec{p} = -p\cdot\hat{e}_z\,(p>0)$ and $\theta_{\text{in}} = 0^{\circ}$. 
In addition, we restricted the incoming muon spectrum to various angular ranges with respect to $\theta_{\text{in}}$.
In particular, we considered angular distributions of 0$^{\circ}$ to 1$^{\circ}$, 0$^{\circ}$ to 5$^{\circ}$ and 0$^{\circ}$ to 25$^{\circ}$.\newline
Simulations were performed for two geometries of the ge{\-}neric model, which differed with respect to the occupancy of the fuel compartments with fuel assemblies.
In the first case, all fuel compartments were empty (no fuel assemblies).
In the second case, all fuel compartments with the exception of the central one were occupied with fuel assemblies.\newline
Projection images were generated using the transmission radiographic as well as the scattering radiographic approach.\linebreak
In case of the transmission radiographic analysis, the major imaging information is the ratio of muons reaching the outgoing detector over the number of muons crossing a specific pixel of the incoming detector.
The pixel size was specified as $(1 \times 1)$\,cm$^2$.\newline
In case of the scattering radiographic analysis, the leading imaging information is the effective scattering angle $\theta_{\text{eff}}$ with respect to the initial direction, which is calculated event-by-event according to:
\begin{equation}
\theta_{\text{eff}} = \arctan\left.\left({ \frac{\sqrt{\sum_{i=x,y}{\left(\Delta_i + \Delta_z \cdot \frac{d_i}{d_z} \right)^2}}}{z_{\text{in}}-z_{\text{out}} }}\right.\right)
\label{Eq:Theta}
\end{equation}
The basis is given by the position information provided by the incoming ($x_{\text{in}}$, $y_{\text{in}}$, $z_{\text{in}}$) and outgoing ($x_{\text{out}}$, $y_{\text{out}}$, $z_{\text{out}}$) detectors as well as the normalized muon direction $\vec{d}=(d_x,d_y,d_z)$ provided by the incoming detector. 
The projected distances $\Delta_i$ ($i=x,y,z$) are given by $i_{\text{out}} - i_{\text{in}}$.\\
Information on positions and directions was processed without any attempts to mimic resolution effects.
Again, the pixel size was defined as $(1 \times 1)$\,cm$^2$.
\subsubsection*{Transmission Radiographic Images}
Fig.\,\ref{Fig:IntView_RadPlots} shows transmission radiographic images for the two geometries and the different angular distributions with respect to $\theta_{\text{in}}$ as well as associated difference images.\\
It can be easily seen that the image quality in terms of resolution decreases with increasing angular acceptance with respect to $\theta_{\text{in}}$.
A good indicator is given in terms of the absorber rods that can be easily identified for the angular acceptance of $0^{\circ}\leq \theta_{\text{in}} \leq 1^{\circ}$.
This, however, is not possible for an angular acceptance of $0^{\circ}\leq \theta_{\text{in}} \leq 5^{\circ}$, at least not with the number of simulated events.\linebreak
A similar picture can be drawn for larger structures such as fuel assemblies: 
For the angular acceptance of $0^{\circ}\leq \theta_{\text{in}} \leq 25^{\circ}$, the transmission radiographic image based on the simulated statistics does not allow for a reliable conclusion on the occupancy of the central fuel compartment with a fuel assembly.\\
It is obvious that difference images between the two geometries benefit from an improved contrast that allows identifying structural differences (or similarities) more easily.
This, for instance, holds with respect to the occupancy of the central fuel compartment.
Even for the angular acceptance of $0^{\circ}\leq \theta_{\text{in}} \leq 25^{\circ}$ the difference image provides weak evidence allowing the conclusion that for both geometries the occupancies of the central fuel compartment were identical.
\subsubsection*{Scattering Radiographic Images}
Fig.\,\ref{Fig:IntView_TomPlots} shows scattering radiographic images for the two  geometries and the various angular acceptances as well as corresponding difference images.\\
Similar to the transmission radiographic images, the image quality deteriorates significantly with the increasing angular acceptance with respect to $\theta_{\text{in}}$.
However, the image quality is much better and smaller structures can be resolved.
For example, the individual walls of the fuel compartments can be easily identified for the angular acceptance of $0^{\circ}\leq \theta_{\text{in}} \leq 1^{\circ}$ and - with limitations - also for the angular acceptance of $0^{\circ}\leq \theta_{\text{in}} \leq 5^{\circ}$.
Unfortunately, also the scattering radiographic images would not allow for a reliable statement on the occupancy of the central fuel compartment for an angular acceptance of $0^{\circ}\leq \theta_{\text{in}} \leq 25^{\circ}$ based on the number of simulated events.\newline
The improved resolution compared to the transmission radiographic images is also reflected in the difference images.
With respect to the angular acceptance of $0^{\circ}\leq \theta_{\text{in}} \leq 25^{\circ}$, at least the difference plot provides clear evidence that the occupancies of the central fuel compartment are identical for both geometries.
\subsubsection*{General Aspects}
The blurring of the structures with increasing angular acceptance can be easily understood in terms of the longitudinal extension of the generic model.\newline
In general, the simulated results show the advantages of the scattering radiographic over the transmission radiographic approach.
Here, the improved resolving power is the most prominent indicator.
Unlike most radiographic detectors, tomographic detection systems would be able to apply cut conditions based on the incoming muon flight directions which would allow us to realize different angular acceptances with respect to $\theta_{\text{in}}$.\newline
In summary, the simulations provide evidence that even without any reconstruction efforts to generate tomographic images it would be possible to make conclusions on the occupancy of specific fuel compartments with fuel assemblies within a reasonable amount of time.
For example, in case of an angular acceptance of $0^{\circ}\leq \theta_{\text{in}} \leq 25^{\circ}$ and based on reasonable assumptions (see Sec.\,\ref{SSec:AnaSingleFuelElement}), the number of simulated events would correspond to a measuring time of about 40\,hours. 
This increases to $\sim 190$\,hours and $\sim 930$\,hours in case of $0^{\circ}\leq \theta_{\text{in}} \leq 5^{\circ}$ and $0^{\circ}\leq \theta_{\text{in}} \leq 1^{\circ}$, respectively.\newline
So far, these results cannot be extrapolated to smaller structures such as individual fuel rods.
Qualitatively, it can be expected that for such a level of detail a narrow angular acceptance as well as a much larger muon flux per area would be required.
The following section provides a more detailed discussion of this aspect.
\subsection{Detailed Longitudinal Study of the Central Fuel Assembly to Detect Missing Fuel Rods}
\label{SSec:AnaSingleFuelElement}
\begin{figure}[b]
\resizebox{1.0\columnwidth}{!}{\includegraphics{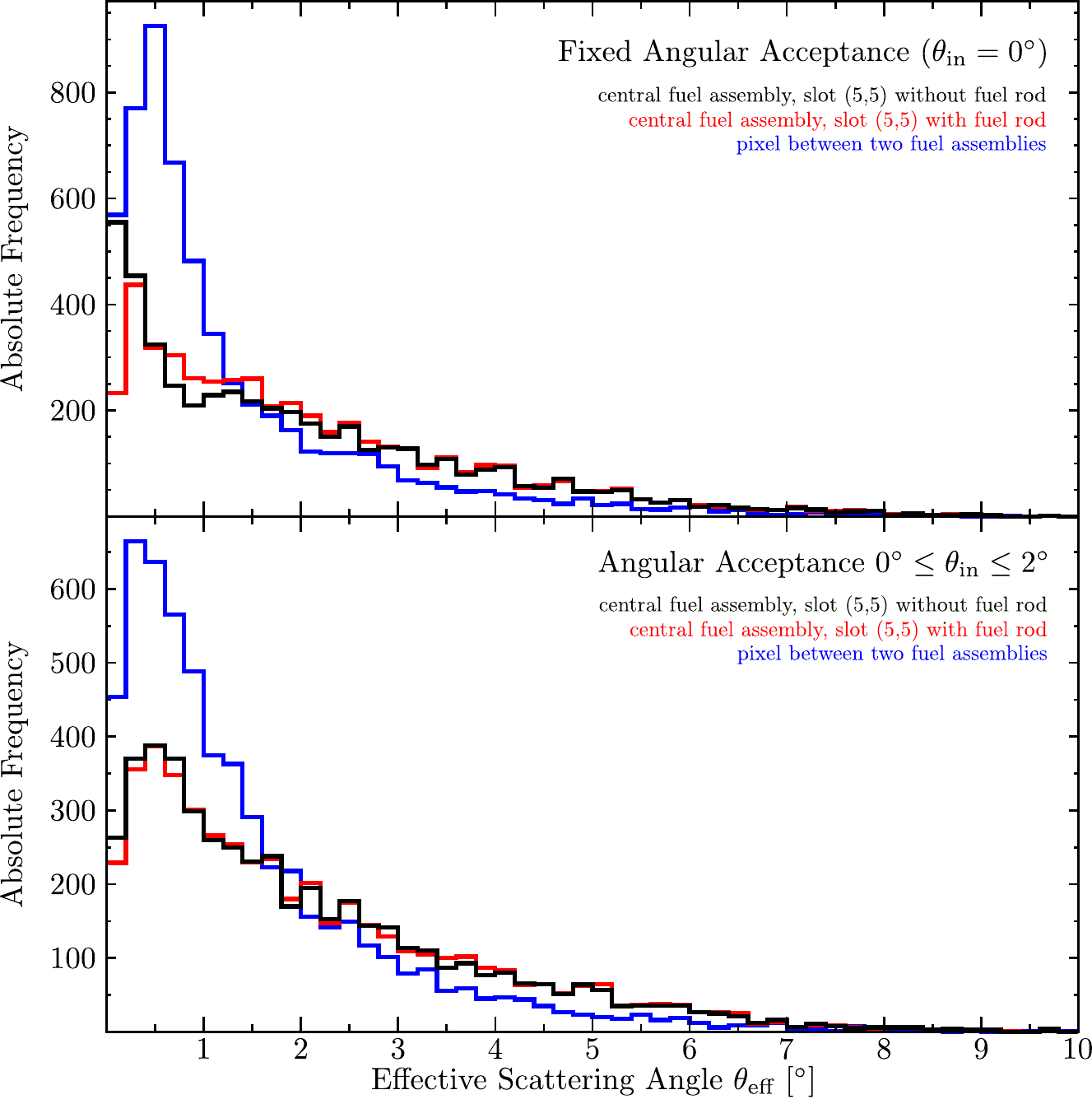}}
\caption{Absolute frequency distributions of the effective scattering angle $\theta_{\text{eff}}$ for different pixels and different angular acceptances. The red and black curves correspond to the pixel of the slot position (5,5) of the central fuel assembly. The black curve represents the case where this particular slot position is empty while the red curve corresponds to the case where the slot position is filled with a fuel rod. The blue curve corresponds to a pixel between two neighbouring fuel assemblies. The \textit{top figure} shows results for a fixed angular acceptance ($\theta_{\text{in}} = 0^{\circ}$) while the \textit{bottom figure} shows results for an angular acceptance  of $0^{\circ}\leq \theta_{\text{in}} \leq 2^{\circ}$. All distributions correspond to simulations with $10^{8}$ events.}
\label{Fig:FrequencyDistrSamples}
\end{figure}

\begin{figure}[t!]
\resizebox{1.0\columnwidth}{!}{\includegraphics{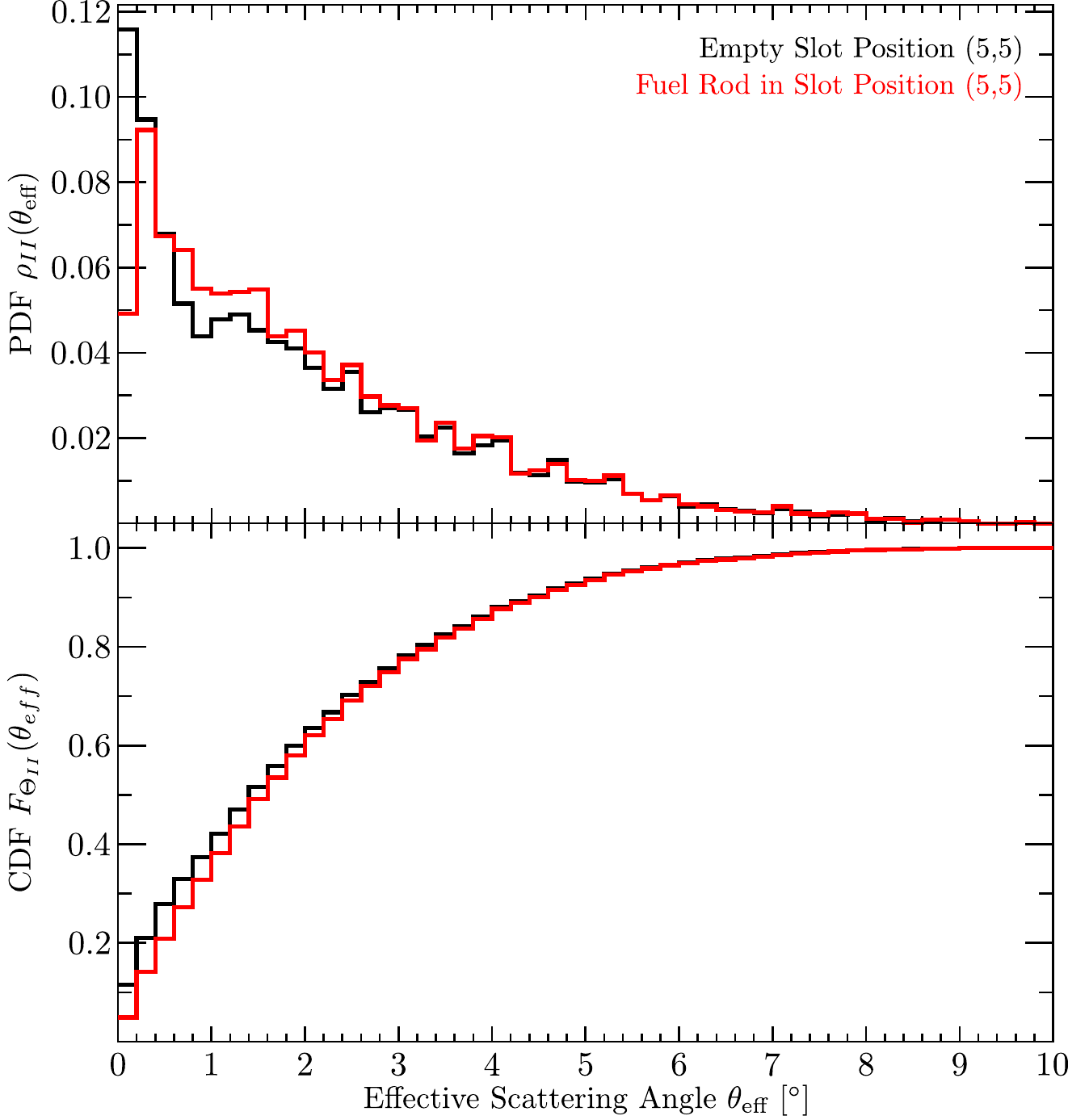}}
\caption{Probability density functions $\rho_{II}(\theta)$ (\textit{top}) as well as cumulative probability functions $F_{\Theta_{II}}(\theta)$ (\textit{bottom}) for the pixel corresponding to the slot position with $x_{\text{id}} = y_{\text{id}} = 5$ within the central fuel assembly for two different geometries. The red curve represents the geometry with a complete fuel assembly (all positions are occupied according to the nominative layout), while the black curve corresponds to the geometry where the specific slot position represented by this pixel is empty. The distributions represent results of a simulation with $10^8$ events assuming mono-directional muons with $\theta_{\text{in}} = 0^{\circ}$.}
\label{Fig:DensFunctions}
\end{figure}

\begin{figure}[t]
\resizebox{1.0\columnwidth}{!}{\includegraphics{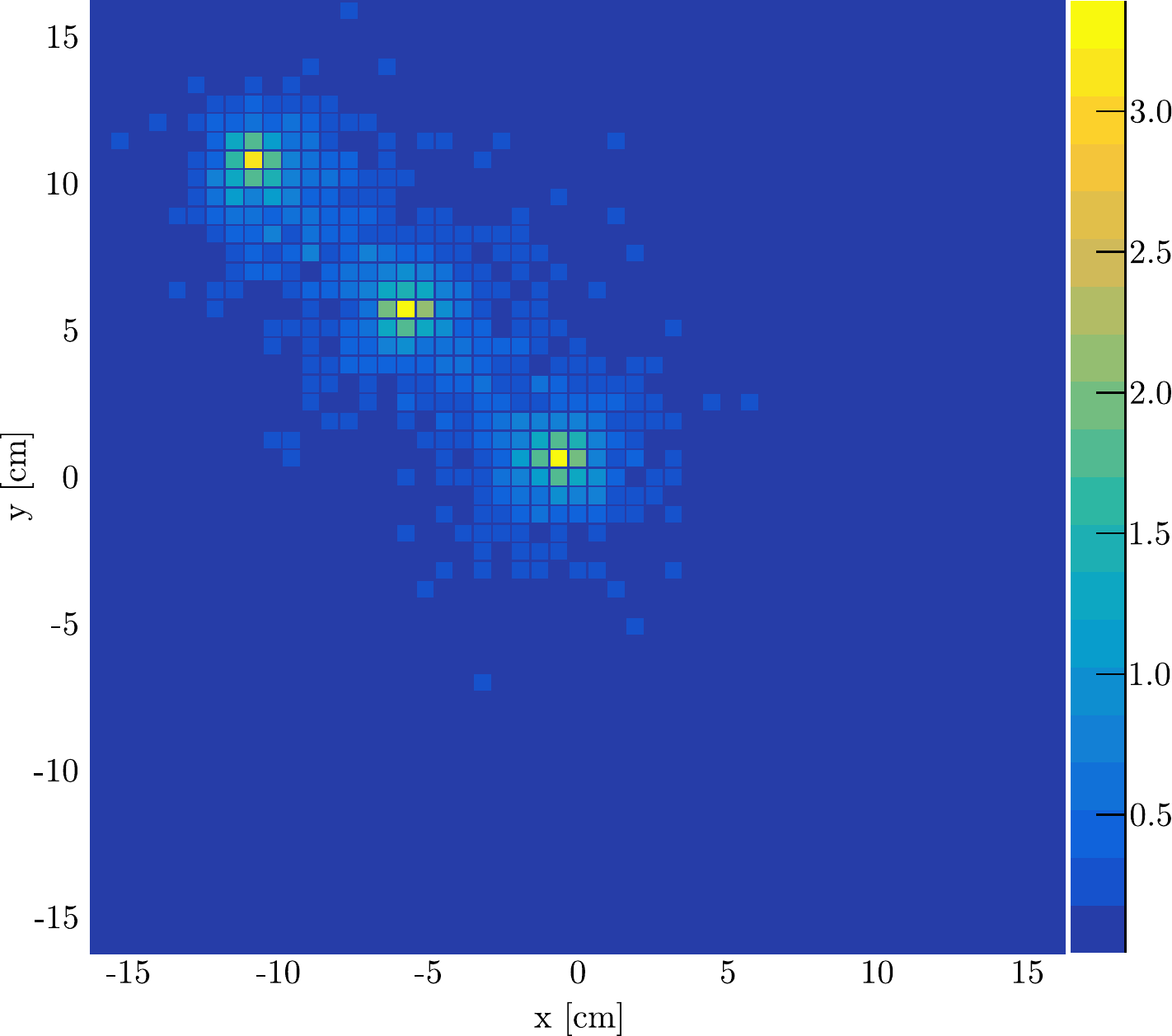}}
\caption{A heat map showing the significance ratio $\tilde{D}$ in the $(x,y)$-plane. One can easily identify the locations of the three vacated fuel-rod slot positions along the diagonal. The spectrum represents simulations with $10^8$ events assuming mono-directional muons. See text for details.}
\label{Fig:DTilde}
\end{figure}

\begin{figure}[t]
\resizebox{1.0\columnwidth}{!}{\includegraphics{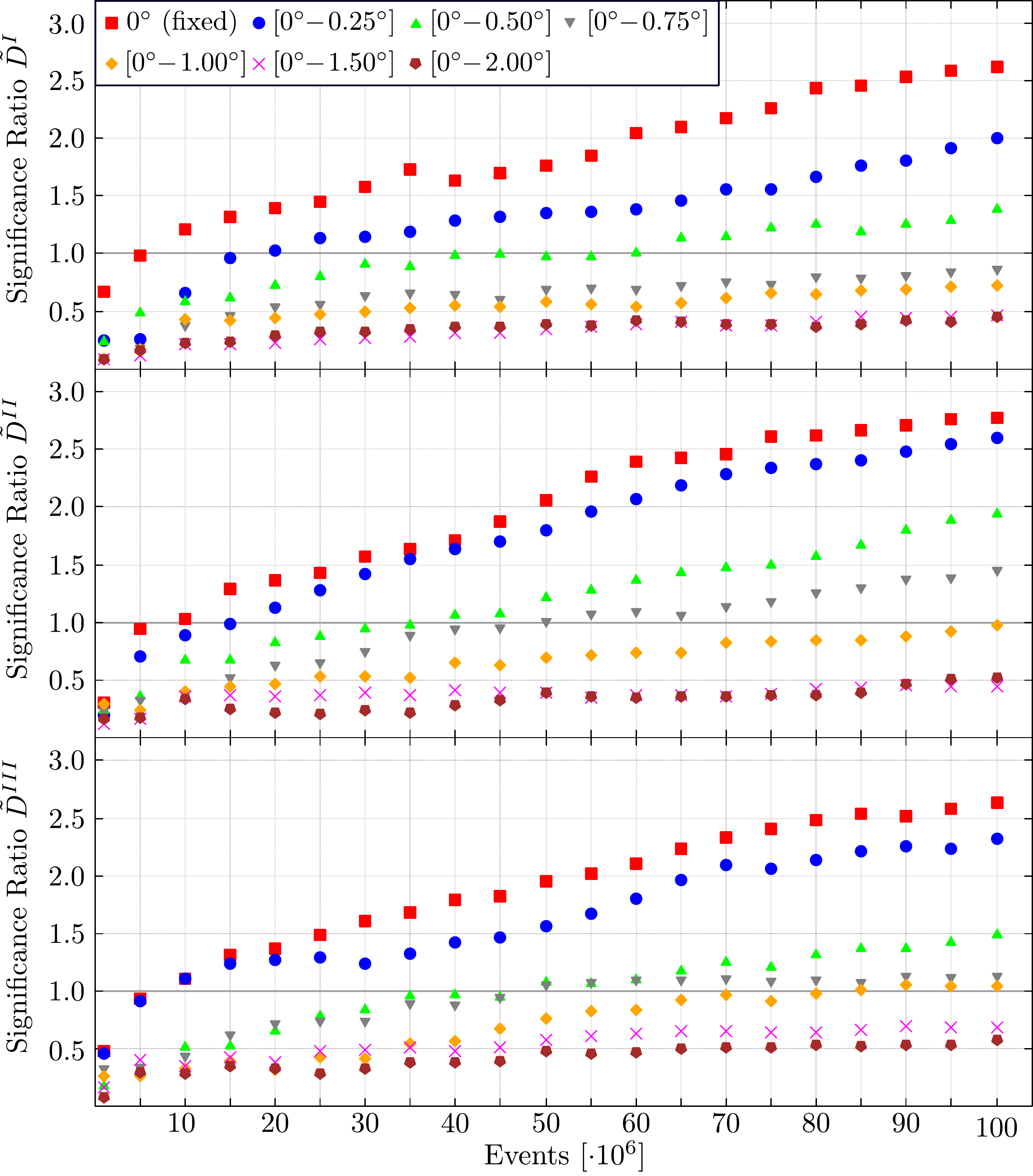}}
\caption{Significance ratios $\tilde{D}^{i,j}$ as a function of angular acceptance with respect to $\theta_{\text{in}}$ and simulated events for the three pixels of interest. The \textit{top figure} shows the results of $D^{I}$, while $D^{II}$ ($D^{III}$) is shown in the \textit{middle} (\textit{bottom}) \textit{figure}.}
\label{Fig:SignificanceLevels}
\end{figure}

In this section, we investigate if muon scattering radiography can be used to make reliable statements about the completeness of individual fuel assemblies.
In particular, we investigate whether individual missing fuel rods in an otherwise complete fuel assembly can be detected.
This investigation takes into account a few variable boundary conditions such as the angular acceptance with respect to $\theta_{\text{in}}$ as well as the number of events, which most often can be used to provide reasonable estimates on the required irradiation time.\newline
\begin{figure}[b]
\resizebox{1.0\columnwidth}{!}{\includegraphics{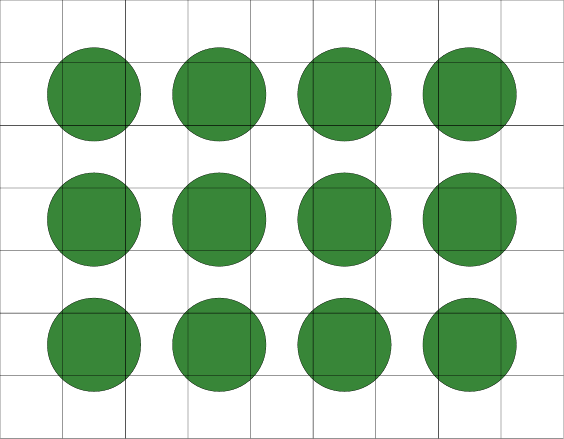}}
\caption{Illustration of the binning used for the analysis in Sec.\,\ref{SSec:AnaSingleFuelElement}. The green disks represent the cross sections of the various rods in the central fuel assembly. The grid indicates the limits of the bins within the $(x,y)$-plane.}
\label{Fig:IntView}
\end{figure}
All simulations were performed with up to $10^8$ events.
The treatment of the absolute muon momenta was identical to the former section \ref{Ssec:OverallScan}.
The considered angular acceptances with respect to $\theta_{\text{in}}$ of the primary muons ranged from $0^{\circ}\leq \theta_{\text{in}} \leq 0.25^{\circ}$ to $0^{\circ}\leq \theta_{\text{in}} \leq 2^{\circ}$.
As in the former section, we also considered mono-directional incoming muons with $\theta_{\text{in}} = 0^{\circ}$ to provide a reference scenario.
The initial positions within the $x,y$-plane were restricted to $-0.35\,\text{m} \leq x,y \leq 0.35\,\text{m}$ and the initial $z$-coordinate was fixed at $z\approx 6.1$\,m.\newline
We performed simulations for two different geometries of the generic model.
The first geometry ensured that all fuel compartments of the model were occupied with fuel assemblies and for each fuel assembly all individual slot positions were occupied according to the nominative layout, see Fig.\,\ref{Fig:RodArrangement}.
The second geometry differs from the first one only by three vacated fuel-rod positions along the diagonal (first, fifth and ninth position - see Fig.\,\ref{Fig:RodArrangement}) within the central fuel assembly.
The first relevant fuel rod $I$ ($x_{\text{id}} = y_{\text{id}} = 1$) is nominatively placed in the upper left corner of the fuel assembly and is surrounded by three fuel rods and the walls of the fuel compartment.
The second relevant fuel rod $II$ ($x_{\text{id}} = y_{\text{id}} = 5$) is surrounded by two control rods and six fuel rods, while the third relevant fuel rod $III$ ($x_{\text{id}} = y_{\text{id}} = 9$) is surrounded by fuel rods on all sides.\newline
Scattering radiographic images were generated using a binning as indicated in Fig.\,\ref{Fig:IntView}.
Each bin covers $(6.36 \times 6.36)\,\text{mm}^2$ and is shifted in a way that for the central fuel assembly the center positions of the bins coincide with the nominative center positions within the $(x,y)$-plane of the various fuel and control rods.\newline
As in the last section, the effective scattering angle $\theta_{\text{eff}}$ is calculated event by event according to Eq.\,\ref{Eq:Theta}.
For each pixel we obtain a histogram that shows the absolute frequencies of the effective scattering angles $\theta_{\text{eff}}$.
Representative examples of such absolute frequency distributions are shown in Fig.\,\ref{Fig:FrequencyDistrSamples}.\newline
These frequency distributions are then normalized for each pixel $(i,j)$ into a normalized probability distribution function (PDF) $\rho_{i,j}(\theta_{\text{eff}})$:
\begin{equation*}
\int_{0}^{\pi/2}{d\theta'_{\text{eff}} \rho_{i,j}(\theta'_{\text{eff}})} = 1
\end{equation*}
In a final step, we use $\rho_{i,j}(\theta_{\text{eff}})$ to calculate for each pixel $i,j$ the cumulative distribution function (CDF) $F_{\Theta_{i,j}}(\theta_{\text{eff}})$, i.e.:
\begin{equation*}
F_{\Theta_{i,j}}(\theta_{\text{eff}}) = \int_{0}^{\theta}{d\theta'_{\text{eff}}\,\rho_{i,j}(\theta'_{\text{eff}})}
\end{equation*}
This procedure is repeated for each simulation.\newline
For illustration, Fig.\,\ref{Fig:DensFunctions} shows for both geometries the PDF $\rho_{II}$ as well as the CDF $F_{\Theta_{II}}$ corresponding to the pixel of the slot position with $x_{\text{id}} = y_{\text{id}} = 5$ within the central fuel assembly.
The analysis focuses on a pixel-based comparison between the two geometries described above.
By computing the local distances between the empirical CDFs, the derived statistical measure can be used for the pair-wise comparison of images \cite{Demidenko2004,Tang2011}.
In other words, we compare for each pixel $i,j$ the CDFs $F^{\text{occ}}_{\Theta_{i,j}}(\theta_{\text{eff}})$ and $F^{\text{vac}}_{\Theta_{i,j}}(\theta_{\text{eff}})$.
Here, $F^{\text{occ}}_{\Theta_{i,j}}(\theta_{\text{eff}})$ corresponds to the geometry for which all slot positions within the fuel assemblies are occupied, while $F^{\text{vac}}_{\Theta_{i,j}}(\theta_{\text{eff}})$ corresponds to the geometry for which the fuel-rod slot positions within the central fuel assembly are empty.\newline 
We consider for each pixel ($i,j$) in the $(x,y)$-plane the two-sample Kolmogorov-Smirnov test,
\begin{equation}
D^{i,j} = \sup_{\theta_{\text{eff}}} \left.\vert{F^{\text{occ}}_{\Theta_{i,j}}(\theta_{\text{eff}}) - F^{\text{vac}}_{\Theta_{i,j}}(\theta_{\text{eff}})}\right.\rvert,
\end{equation}
which allows us to make quantitative statements on the agreement between $F^{\text{occ}}_{\Theta_{i,j}}(\theta_{\text{eff}})$ and $F^{\text{vac}}_{\Theta_{i,j}}(\theta_{\text{eff}})$.\newline
The null hypothesis - i.e. $F^{\text{occ}}_{\Theta_{i,j}}(\theta_{\text{eff}})$ and $F^{\text{vac}}_{\Theta_{i,j}}(\theta_{\text{eff}})$ describe identical distributions - is rejected at a statistical significance level $\alpha$ if the test statistic satisfies
\begin{eqnarray*}
D^{i,j} &>& c(\alpha)\sqrt{\frac{n_{i,j} + m_{i,j}}{n_{i,j}\cdot m_{i,j}}}\\
\label{Eq:D}
c(\alpha) &=& \sqrt{-0.5\cdot\ln\left(\alpha/2 \right)}
\end{eqnarray*}
Here $n_{i,j}$ ($m_{i,j}$) describes the number of entries in the $(i,j)$ pixel of the relevant vacated (occupied) spectrum. 
The condition can be reformulated as:
\begin{equation}
\tilde{D}^{i,j} \equiv \frac{D^{i,j}}{c(\alpha)} \sqrt{\frac{n_{i,j}\cdot m_{i,j}}{n_{i,j} + m_{i,j}}} > 1
\end{equation}
We specified the significance level $\alpha$ of 0.1 in the following.\newline
Fig.\,\ref{Fig:DTilde} shows a heatmap of $\tilde{D}^{i,j}$ based on a simulation with $10^8$ events assuming mono-directional muons with $\theta_{\text{in}} = 0^{\circ}$.
One can clearly identify areas for which the Kolmogorov-Smirnov test statistic indicates the significant deviations.
These areas coincide perfectly with areas for which the two geometries differ, i.e. with respect to the occupation of the slot positions $x_{\text{id}} = y_{\text{id}} = 1$ ($I$), $x_{\text{id}} = y_{\text{id}} = 5$ ($II$) and $x_{\text{id}} = y_{\text{id}} = 9$ ($III$) within the central fuel assembly.\newline
Fig.\,\ref{Fig:SignificanceLevels} summarizes $\tilde{D}^{i,j}$ for each of these three particular pixels and for different angular acceptances and quantifies the evolution as a function of simulated events.
$\tilde{D}^{I}$ corresponds to the pixel of $x_{\text{id}} = y_{\text{id}} = 1$, while $\tilde{D}^{II}$ and $\tilde{D}^{III}$ correspond to the pixels of $x_{\text{id}} = y_{\text{id}} = 5$ and $x_{\text{id}} = y_{\text{id}} = 9$. 
\newline
It is obvious that for all three pixels the significance ratios $\tilde{D}$ decrease in general with an increasing angular acceptance.
The only exception is given in terms of the angular acceptances $0^{\circ} \leq \theta_{\text{in}} \leq 1.5^{\circ}$ and $0^{\circ} \leq \theta_{\text{in}} \leq 2^{\circ}$, for which comparable significance ratios are observed.
For all three pixels one can observe a clear trend towards smaller slopes of progression with an increasing angular acceptance and increasing number of simulated events.
It is interesting to note that there are significant differences between the three pixels with respect to the trends with increasing event numbers and the achieved significance ratios.
These effects are stronger for pixel $II$ ($III$) compared to $I$.
With respect to pixel $I$ and based on the maximum number of events ($10^8$), the significant ratio $\tilde{D}^{I}$ exceeds one only up to an angular acceptance of $0^{\circ} \leq \theta_{\text{in}} \leq 0.5^{\circ}$ and requires at least $6\cdot 10^7$ simulated events in case of the latter.
A different picture can be drawn for pixel $II$.
Not only do we observe that less events ($4\cdot 10^7$ events) are required for $\tilde{D}^{II}$ to exceed one for $0^{\circ} \leq \theta_{\text{in}} \leq 0.5^{\circ}$, we also observe significant deviations with respect to $0^{\circ} \leq \theta_{\text{in}} \leq 0.75^{\circ}$.
For the latter, at least $5\cdot 10^7$ simulations events are required for $\tilde{D}^{II}$ to exceed one.
In case of pixel $III$, we observe that $\tilde{D}^{III}$ also exceeds one  for $0^{\circ} \leq \theta_{\text{in}} \leq 1.0^{\circ}$, requiring at least $8.5\cdot 10^7$ simulated events. 
The significant deviations between the individual pixels may be related to the larger numbers of neighbouring fuel rods in case of pixels $II$ and $III$ compared to pixel $I$.\newline
For a specific simulation with known momentum distribution and angular acceptance, the measurement time can be estimated according to the following formula:
\begin{equation}
\Delta t [\text{s}] = \frac{\text{simulated events}}{A[\text{cm}^2]\cdot I_{\mu}[\text{muons}/\text{cm}^2/\text{s}]\cdot I_{\theta_{\text{in}}} \cdot I_{p}\cdot \epsilon }
\label{Eq:TimeEstimate}
\end{equation}
Here, $A$ is the area of the initial $(x,y)$-plane, $I_{\mu}$ is the integrated muon flux at sea level ($\sim 1\,\text{muon}/\text{cm}^2/\text{min}$), $I_{\theta_{\text{in}}}$ is the share within the angular acceptance with respect to $\theta_{\text{in}}$, $I_{p}$ is the share of the considered momentum distribution with respect to the full momentum distribution and $\epsilon$ is the efficiency of the detector system.
For a certain angular acceptance, $I_{\theta_{\text{in}}}$ can be estimated according to:
\begin{equation}
I_{\theta_{\text{in}}} = \frac{\int_{0}^{\theta_{\text{in}}}{d\theta ' \cos^2\theta '}}{\int_{0}^{\pi/2}{d\theta ' \cos^2\theta '}}
\end{equation}
Values of $I_{\theta_{\text{in}}}$ for various angular acceptances are listed in Tab.\,\ref{Tab:ITheta}.
In a similar way, a simplified estimate of $I_{p}$ ($\approx 0.8$) can be determined by means of Eq.\,\ref{Eq:Reyna}:
\begin{equation}
I_{p} = \frac{\int_{p(T=1\,\text{GeV})}^{p(T=1\,\text{TeV})}{dp ' I(p'_{\mu}, 0)}}{\int_{0}^{\infty}{dp ' I(p'_{\mu}, 0)}}
\end{equation}
With respect to $0^{\circ} \leq \theta_{\text{in}} \leq 0.5^{\circ}$, $\tilde{D}^I$ exceeds 1.0 for $6\cdot 10^7$ events.
Based on the above estimates, this event number corresponds to a measurement time of $\sim 2.5$\,years.
Slightly fewer events ($4\cdot 10^7$ and $5\cdot 10^7$) are required for the pixels II and III, which leads to estimated measurement times of $\sim 1.7$\,years and $\sim 2.1$\,years.
\begin{table}[b]
\begin{center}
\caption{Estimations for $I_{\theta_{\text{in}}}$ for various angular acceptances.}
\label{Tab:ITheta}       
\begin{tabular}{lccccc}
\hline\noalign{\smallskip}
 & [$0^{\circ},0.5^{\circ}$] & [$0^{\circ},1^{\circ}$] & [$0^{\circ},2^{\circ}$] & [$0^{\circ},5^{\circ}$] & [$0^{\circ},25^{\circ}$]\\
\noalign{\smallskip}\hline\noalign{\smallskip}
$I_{\theta_{\text{in}}}$ [\%] &  1.11 & 2.22 & 4.44 & 11.08 & 52.16\\
\noalign{\smallskip}\hline
\end{tabular}
\end{center}
\end{table}
%
\section{Summary and conclusion}
\label{sec:Summary}
In this work we investigated the theoretical feasibility to detect individual missing fuel rods in an otherwise fully loaded cask within a reasonable amount of time using cosmic muons. 
We used simulations based on the \textsc{Geant4} toolkit and a generic model based on the CASTOR\textsuperscript{\textregistered} V/19 (see Fig.\,\ref{Fig:GenModelExplodedView}) loaded with 18x18-24 PWR fuel assemblies (see Fig.\,\ref{Fig:RodArrangement}).
The detectors above and below a vertical (standing) model were mimicked by two rectangular planes with vanishing thickness and an area of 9 m$^2$. 
The gap between detectors and the model was assumed to be 10 cm. 
The muon characteristics are described in section \ref{ssec:PrimaryParticles}. 
The tool can be run with a realistic muon spectrum but also allows for performing simulations assuming mono-energetic and mono-directional muons. 
For validation purposes, the stopping powers (see Fig.\,\ref{Fig:ValdEdx}) and scattering angles (see Fig.\,\ref{Fig:ValScattering}) of the relevant target materials have been investigated and compared against established reference values from the literature. 
We found good agreement between simulated and literature/analytical values.\newline
We simulated longitudinal scans that covered the full cross section of the model (full-model simulation). 
In addition, we put a special emphasis on a single fuel assembly with missing fuel rods.\newline
The focus of the full-model simulation was on the effect of angular acceptance criteria on the resolution.
We compared mono-directional muons ($\theta_{\text{in}} = 0^{\circ}$) with different angular acceptance ranges with respect to $\theta_{\text{in}}$: $0^{\circ} \leq \theta_{\text{in}} \leq 1^{\circ}$, $0^{\circ} \leq \theta_{\text{in}} \leq 5^{\circ}$ and $0^{\circ} \leq \theta_{\text{in}} \leq 25^{\circ}$. 
The integrated flux was kept constant with 5$\cdot$10$^{7}$ muons, which corresponds to a radiation time of approximately 40 hours in case of $0^{\circ} \leq \theta_{\text{in}} \leq 25^{\circ}$.
We performed two analyses and compared the empty model with a partially loaded one and the resulting difference between the two radiographic images.
The partially loaded model was lacking the central fuel assembly. 
In the first analysis (transmission radiographic scans) we compared the ratio of muons in the lower and upper detectors. 
This radiographic analysis clearly indicated the decreasing resolution with increasing angular acceptance with respect to $\theta_{\text{in}}$ of the incoming muons (see Fig.\,\ref{Fig:IntView_RadPlots}). 
%
%
The central fuel assembly could still be identified as missing based on the difference plot of the acceptance criteria $0^{\circ} \leq \theta_{\text{in}} \leq 25^{\circ}$.
In addition, we repeated the analysis with a focus on the effective scattering angles of the muons to receive scattering radiographic images. 
The image quality was much better compared to the transmission radiographic  analysis and more details could be identified (see Fig.\,\ref{Fig:IntView_TomPlots}). 
Again, for the acceptance criteria of $0^{\circ} \leq \theta_{\text{in}} \leq 25^{\circ}$ the difference plot can be used to identify the central fuel assembly as missing.\newline
We further investigated the capability of muon scattering radiography to provide reliable statements on the completeness of a loaded fuel assembly, assuming up to $10^8$ events and angular acceptance criteria of $0^{\circ} \leq \theta_{\text{in}} \leq 0.25^{\circ}$, $0^{\circ} \leq \theta_{\text{in}} \leq 0.5^{\circ}$, $0^{\circ} \leq \theta_{\text{in}} \leq 0.75^{\circ}$, $0^{\circ} \leq \theta_{\text{in}} \leq 1.0^{\circ}$, $0^{\circ} \leq \theta_{\text{in}} \leq 1.5^{\circ}$ and $0^{\circ} \leq \theta_{\text{in}} \leq 2.0^{\circ}$ in addition to mono-directional muons with $\theta_{\text{in}}=0^{\circ}$.
Focussing on the central fuel assembly we investigated a fully loaded model and one where the central assembly was missing three fuel rods on the diagonal (see Fig.\,\ref{Fig:RodArrangement}).
All three missing rods were clearly identifiable after the simulation of $10^8$ events assuming mono-directional muons with $\theta_{\text{in}}=0^{\circ}$ (see Fig.\,\ref{Fig:DTilde}).
The significance ratio $\tilde{D}$ based on the Kolmogorov-Smirnov test statistic of each fuel rod (and thus the significance of the contrast of a missing rod) depends on the relative position of the missing fuel rod in the assembly, the angular acceptance criteria of incoming muons and the number of simulated events (see Fig.\,\ref{Fig:SignificanceLevels}).
The latter can be transformed accordingly into radiation or measuring times.
Assuming significance ratios $\tilde{D}^{i,j}>1$ leading to a successful identification of missing fuel rods we found measuring times of approximately 2 years for the two inner missing rods and 2.5 years for the corner rod assuming an angular acceptance of $0^{\circ} \leq \theta_{\text{in}} \leq 0.5^{\circ}$.\newline
In summary, we have shown that the muon scattering radiography is capable of reliably visualizing the inside of a loaded and sealed model at a resolution scale of a single fuel rod. 
The assumptions we made led to timescales which seem feasible compared to the duration of the dry storage of spent nuclear fuel. 
In the next section we discuss how future work might even shorten the theoretically estimated measuring time and how image processing methods may be used to gain more detailed insights into the cask's interior for the detection of individual missing fuel rods.
\section{Outlook}
\label{sec:Outlook}
The present work will serve as a starting point for further research activities that may evolve into several independent aspects.\newline
The first aspect relates to the developed simulation tool itself.
We intend to include the discussion of transversal scans, for which the detectors are located on the sides of the generic model.
The simulation can also be extended to additional geometries which do not have to be limited to storage casks.\newline
The second aspect relates to the validation of the simulated data and comparisons to other simulations.
We already addressed validation aspects in the present work that concerned the slowing-down process of the muons as well as the angular scattering.
A logical next step would be given by a comparison of simulated to experimental results.
For this it might be reasonable to start with less complex geometries or larger objects-of-interest which will require less time, both with respect to the computation and the experimental measurement.
To take into account the uncertainty due to a particular choice of the simulation tool, it would be valuable to compare the results from different simulation tools for identical geometries for benchmark purposes.\newline
A third aspect relates to sensitivity analyses.
The present results indicate that the significance depends on the relative fuel rod position and, hence, the immediate surroundings of the considered rod. 
So far, we have arbitrarily selected three fuel rods and it may be worth the effort to repeat the analyses systematically, including the effects of rods in the immediate surroundings.
The significance information of the second analysis part is derived from two predominantly identical geometries, which deviated only by the occupation of three fuel rod slots. 
It would be insightful to investigate the effects of additional differences such as slight misalignments.\newline
A fourth aspect addresses the extension to tomographic images which allows using the complete information of scattered (or absorbed) muons within one visualization processed by various potential statistical muon tomography reconstruction algorithms as discussed in \cite{Riggi2013}. 
For adaptively comparing muon scattering (or absorption) images of a cask with the purpose of automatically detecting changes within the interior of the cask, the image processing strategy needs to be extended towards tomographic image reconstruction inherently designed for change detection or in combination with further image analysis methods.  
\section{Acknowledgements}
This research was partially funded by the Federal Ministry for the Environment, Nature Conservation and Nuclear Safety in Germany (BMU) under Contract 4720E03366.
%
%


\begin{thebibliography}{}
%
%
\bibitem{Zio05}
K.-P. Ziock, G. Caffrey, A. Lebrun, L. Forman, P. Vanier and J. Wharton, \textit{IEEE Nuclear Science Symposium Conference Record, 2005, Fajardo, 2005, pp. 1163-1167}, doi:\href{https://doi.org/10.1109/NSSMIC.2005.1596457}{10.1109/NSSMIC.2005.1596457}

\bibitem{Brdar:2017}
V. Brdar, P. Huber and J. Kopp, Phys. Rev. Appl. \textbf{8}, 2331 (2017), doi:\href{https://doi.org/10.1103/PhysRevApplied.8.054050}{10.1103/PhysRevApplied.8.054050}
%
\bibitem{Bon20}
G. Bonomi, P. Checchia, M. D'Errico, D. Pagano and G. Saracino, Prog. Part. Nucl. Phys. \textbf{112}, 103768 (2020), doi:\href{https://doi.org/10.1016/j.ppnp.2020.103768}{10.1016/j.ppnp.2020.103768}
%
\bibitem{RPP20}
P.A. Zyla \textit{et al.} (Particle Data Group), Prog. Theor. Exp. Phys. 2020, 083C01 (2020), doi:\href{https://doi.org/10.1093/ptep/ptaa104}{10.1093/ptep/ptaa104}
%
\bibitem{Pas93} 
M.P. De Pascale \textit{et al.}, J. Geophys. Res. A \textbf{98}, 3501 (1993), doi:\href{https://doi.org/10.1029/92JA02672}{10.1029/92JA02672}
%
\bibitem{Gri01} 
P.K.F. Grieder, \textit{Cosmic Rays at Earth} (Elsevier Science, Amsterdam, 2001)
%
\bibitem{Alv70}
L.W. Alvarez \textit{et al.}, Science \textbf{167}, 832-839 (1970), doi:\href{https://doi.org/10.1126/science.167.3919.832}{10.1126/science.167.3919.832}
%
\bibitem{Tan14}
H.K.M. Tanaka, T. Kusagaya and H. Shinohara, Nat. Commun. \textbf{5}, 3381  (2014), doi:\href{https://doi.org/ 10.1038/ncomms4381 (2014)}{10.1038/ncomms4381 (2014)}
%
\bibitem{Les10}
N. Lesparre, D. Gilbert, J. Marteau, Y. Déclais, D. Carbone and E. Galichet, Geophys. J. Int. \textbf{183}, 1348-1361 (2010), doi:\href{https://doi.org/10.1111/j.1365-246X.2010.04790.x}{10.1111/j.1365-246X.2010.04790.x}
%
\bibitem{Mor17}
K. Morishima \textit{et al.}, Nature \textbf{552}, 386-390 (2017), \href{https://doi.org/10.1038/nature24647}{10.1038/nature24647}
%
\bibitem{Tan05}
H.K.M. Tanaka, K. Nagamine, S.N. Nakamura and K. Ishida, Nucl. Instrum. Methods Phys. Res. A \textbf{555}, 164-172 (2005), doi:\href{https://doi.org/10.1016/j.nima.2005.08.099}{10.1016/j.nima.2005.08.099}
%
\bibitem{Bor03}
K. Borozdin \textit{et al.}, Nature \textbf{422}, 277 (2003), doi:\href{https://doi.org/10.1038/422277a}{10.1038/422277a}
%
\bibitem{Cla15}
A. Clarkson \textit{et al.}, J. Instrum. \textbf{10}, P03020 (2015), doi:\href{https://doi.org/10.1088/1748-0221/10/03/P03020}{10.1088/1748-0221/10/03/P03020}
%
\bibitem{Dec21}
Decision Sciences \url{https://www.decisionsciences.com/our-product/}.
%
\bibitem{Zen14}
A. Zenoni. \textit{2015 4th International Conference on Advancements in Nuclear Instrumentation Measurement Methods and their Applications (ANIMMA)} (IEEE, Lisbon, 2015), doi:\href{https://doi.org/10.1109/ANIMMA.2015.7465542}{10.1109/ANIMMA.2015.7465542}
%
\bibitem{Dur18}
J.M. Durham \textit{et al.}, Phys. Rev. Appl. \textbf{9}, 044013 (2018), doi:\href{https://doi.org/10.1103/PhysRevApplied.9.044013}{10.1103/PhysRevApplied.9.044013}
%
\bibitem{Jon13}
G. Jonkmans, V.N.P. Anghel, C. Jewett and M. Thompson, Ann. Nucl. Energy \textbf{53}, 267 (2013), doi:\href{https://doi.org/10.1016/j.anucene.2012.09.011}{10.1016/j.anucene.2012.09.011}
%
\bibitem{Cla14}
A. Clarkson \textit{et al.}, Nucl. Instrum. Methods Phys. Res. A \textbf{746}, 64-73 (2014), doi:\href{https://doi.org/10.1016/j.nima.2014.02.019}{10.1016/j.nima.2014.02.019}
%
\bibitem{Cha14}
S. Chatzidakis, M. Alamaniotis and L. H. Tsoukalas, Trans. Am. Nucl. Soc. \textbf{111}, 369 (2014), \url{https://ans.org/pubs/transactions/a_36379}
%
\bibitem{Amb15}
F. Ambrosino \textit{et al.}, J. Instrum. \textbf{10}, T06005 (2015), doi:\href{https://doi.org/10.1088/1748-0221/10/06/T06005}{10.1088/1748-0221/10/06/T06005}
%
\bibitem{Pou17}
D. Poulson, J.M. Durham, E. Guardincerri, C.L. Morris, J.D. Bacon, K. Plaud-Ramos, D. Morley and A.A. Hecht, Nucl. Instrum. Methods Phys. Res. A \textbf{842}, 48 (2017), doi:\href{https://doi.org/10.1016/j.nima.2016.10.040}{10.1016/j.nima.2016.10.040}
%
\bibitem{Pou18}
D. Poulson, J. Bacon, M. Durham, E. Guardincerri, C. L. Morris and H.R. Trellue, Phil. Trans. Roy. Soc. A \textbf{377}, 2137 (2018), doi:\href{https://doi.org/10.1098/rsta.2018.0052}{10.1098/rsta.2018.0052}
%
\bibitem{Ago03}
S. Agostinelli \textit{et al.}, Nucl. Instrum. Methods Phys. Res. A \textbf{506}, 250-303 (2003), doi:\href{https://doi.org/10.1016/S0168-9002(03)01368-8}{10.1016/S0168-9002(03)01368-8}
%
\bibitem{All06}
J. Allison \textit{et al.}, IEEE Trans. Nucl. Sci. \textbf{53}, 270-278 (2006), doi:\href{https://doi.org/10.1109/TNS.2006.869826}{10.1109/TNS.2006.869826}
%
\bibitem{All16}
J. Allison \textit{et al.}, Nucl. Instrum. Methods Phys. Res. A \textbf{835}, 186(2016), doi:\href{https://doi.org/10.1016/j.nima.2016.06.125}{10.1016/j.nima.2016.06.125}
%
\bibitem{Bru97}
R. Brun and F. Rademakers,  Nucl. Instrum. Methods Phys. Res. A \textbf{389}, 81-86 (1997), doi:\href{https://doi.org/10.1016/S0168-9002(97)00048-X}{10.1016/S0168-9002(97)00048-X}
%
\bibitem{ROOT20}
See also "ROOT" [software], Release v6.20/04, 01/04/2020 
%
\bibitem{GNS20}
GNS Geselllschaft für Nuklear-Serivce mbH, Product Info Castor\textsuperscript{\textregistered} V/19
%
\bibitem{BFS00}
Bundesamt für Strahlenschutz (BfS), \textit{Radioaktive Frachten unterwegs - Atomtransporte und Sicherheit} (brochure, 2000)
%
\bibitem{CMS10}
CMS Collaboration, Phys. Lett. B \textbf{692}, 83 (2010), doi:\href{https://doi.org/10.1016/j.physletb.2010.07.033}{10.1016/j.physletb.2010.07.033}
%
\bibitem{Rey06}
D. Reyna, arXiv:\href{https://arxiv.org/abs/hep-ph/0604145}{hep-ph/0604145v2}
%
\bibitem{Bug98}
E.V. Bugaev, A. Misaki, V.A. Naumov, T. Sinegovskaya, S.I. Sinegovsky, and N. Takahashi, Phys. Rev. D \textbf{58}, 054001 (1998), doi:\href{https://doi.org/10.1103/PhysRevD.58.054001}{10.1103/PhysRevD.58.054001}
%
\bibitem{Eat19}
J.W. Eaton, D. Bateman, S. Hauberg and R. Wehbring, GNU Octave version 5.2.0 manual: a high-level interactive language for numerical computations. \url{https://www.gnu.org/software/octave/doc/v.5.2.0/}
%
\bibitem{Gro01}
D.E. Groom, N.V. Mokhov and S.I. Striganov, At. Data Nucl. Data Tables \textbf{78}, 183 (2001), doi:\href{https://doi.org/10.1006/adnd.2001.0861}{10.1006/adnd.2001.0861}
%
\bibitem{ANP20}
\url{https://pdg.lbl.gov/2020/AtomicNuclearProperties/}
%
\bibitem{Bra05}
W.H. Bragg and R. Kleeman, Philos. Mag. \textbf{10}, 318-340 (1905), doi:\href{https://doi.org/10.1080/14786440509463378}{10.1080/14786440509463378}
%
\bibitem{Lyn91}
G.R. Lynch and O.I. Dahl, Nucl. Instrum. Methods Phys. Res. B \textbf{58}, 6 (1991), doi:\href{https://doi.org/10.1016/0168-583X(91)95671-Y}{10.1016/0168-583X(91)95671-Y}

\bibitem{Demidenko2004}
E. Demidenko, In: Laganá A., Gavrilova M.L., Kumar V., Mun Y., Tan C.J.K., Gervasi O. (eds) Computational Science and Its Applications – ICCSA 2004. ICCSA 2004. Lecture Notes in Computer Science, vol 3046. (Springer-Verlag Berlin, Heidelberg, 2004), doi:\href{https://doi.org/10.1007/978-3-540-24768-5_100}{10.1007/978-3-540-24768-5{\_}100}
%
\bibitem{Tang2011}
Y. Tang, L. Zhang and X. Huang, Int. J. Remote Sens. \textbf{32}, 5719 (2011), doi:\href{https://doi.org/10.1080/01431161.2010.507263}{10.1080/01431161.2010.507263}
%
\bibitem{Riggi2013}
S. Riggi \textit{et. al.}, Nucl. Instrum. Methods Phys. Res. A 728, 59 (2013), doi:\href{https://doi.org/10.1016/j.nima.2013.06.040}{10.1016/j.nima.2013.06.040}
%
\end{thebibliography}
\end{document}